\documentclass[onecolumn]{aastex631}
\usepackage{amsmath,amstext}
\usepackage[T1]{fontenc}
\usepackage{apjfonts} 
\usepackage[figure,figure*]{hypcap}
\usepackage{CJKutf8}
\usepackage{longtable}

\defcitealias{huang2022}{H2022}
\defcitealias{vanLaerhoven2019}{VL2019}

\shortauthors{Pike et al.}

\begin{document}

\title{Col-OSSOS: The Distribution of Surface Classes in Neptune's Resonances}

\author[0000-0003-4797-5262]{Rosemary~E. Pike}
\affiliation{Center for Astrophysics | Harvard \& Smithsonian; 60 Garden Street, Cambridge, MA, 02138, USA}

\author[0000-0001-6680-6558]{Wesley C. Fraser}
\affiliation{National Research Council of Canada, 5071 West Saanich Rd, Victoria, BC, Canada}

\author[0000-0001-8736-236X]{Kathryn Volk}
\affiliation{Lunar and Planetary Laboratory, University of Arizona, 1629 E University Blvd, Tucson, AZ 85721, USA}
\affiliation{Planetary Science Institute, 1700 East Fort Lowell, Suite 106, Tucson, AZ 85719, USA}

\author[0000-0001-7032-5255]{J.J. Kavelaars}
\affiliation{Herzberg Astronomy and Astrophysics Research Centre, National Research Council of Canada, 5071 West Saanich Rd, Victoria, BC, Canada}

\author[0000-0001-8617-2425]{Micha\"el Marsset}
\affiliation{European Southern Observatory (ESO), Alonso de Cordova 3107, 1900 Casilla Vitacura, Santiago, Chile}
\affiliation{Department of Earth, Atmospheric and Planetary Sciences, MIT, 77 Massachusetts Avenue, Cambridge, MA 02139, USA}

\author[0000-0002-6830-476X]{Nuno Peixinho}
\affiliation{Instituto de Astrof\'{\i}sica e Ci\^{e}ncias do Espa\c{c}o, Departamento de F\'{\i}sica, Universidade de Coimbra, 3040-004 Coimbra, Portugal}

\author[0000-0003-4365-1455]{Megan E. Schwamb}
\affiliation{Astrophysics Research Centre, Queen's University Belfast, Belfast BT7 1NN, UK}

\author[0000-0003-3257-4490]{Michele T. Bannister}
\affiliation{School of Physical and Chemical Sciences | Te Kura Mat\=u University of Canterbury, Private Bag 4800, Christchurch 8140, New Zealand}

\author[0000-0002-9179-8323]{Lowell Peltier}
\affil{Department of Physics and Astronomy, University of Victoria, Elliott Building, 3800 Finnerty Rd, Victoria, BC V8P 5C2, Canada}

\author[0000-0002-8032-4528]{Laura E. Buchanan}
\affiliation{Astrophysics Research Centre, Queen's University Belfast, Belfast BT7 1NN, UK}

\author[0000-0001-8821-5927]{Susan Benecchi}
\affiliation{Planetary Science Institute, 1700 East Fort Lowell, Suite 106, Tucson, AZ 85719}

\author[0000-0001-6541-8887]{Nicole J. Tan}
\affiliation{School of Physical and Chemical Sciences | Te Kura Matū, University of Canterbury, Private Bag 4800, Christchurch 8140, New Zealand}

\begin{abstract}

The distribution of surface classes of resonant trans-Neptunian objects (TNOs) provides constraints on the proto-planetesimal disk and giant planet migration. 
To better understand the surfaces of TNOs, the Colours of the Outer Solar System Origins Survey (Col-OSSOS) acquired multi-band photometry of 102 TNOs, and found that the surfaces of TNOs can be well described by two surface classifications, BrightIR and FaintIR.
These classifications both include optically red members and are differentiated predominantly based on whether their near-infrared spectral slope is similar to their optical spectral slope.
The vast majority of cold classical TNOs, with dynamically quiescent orbits, have the FaintIR surface classification, and we infer that TNOs in other dynamical classifications with FaintIR surfaces share a common origin with the cold classical TNOs.
Comparison between the resonant populations and the possible parent populations of cold classical and dynamically excited TNOs reveal that the 3:2 has minimal contributions from the FaintIR class, which could be explained by the $\nu_8$ secular resonance clearing the region near the 3:2 before any sweeping capture occurred.
Conversely, the fraction of FaintIR objects in the 4:3 resonance, 2:1 resonance, and the resonances within the cold classical belt, suggest that the FaintIR surface formed in the protoplanetary disk between $\gtrsim34.6$~au to $\lesssim$47~au, though the outer bound depends on the degree of resonance sweeping during migration.
The presence and absence of the FaintIR surfaces in Neptune's resonances provides critical constraints for the history of Neptune's migration, the evolution of the $\nu_8$, and the surface class distribution in the initial planetesimal disk.

\end{abstract}

\section{Introduction}

The surface colors of Trans-Neptunian Objects (TNOs) in a variety of optical and near-infrared wavelengths have been used to classify TNOs into different surface classifications. 
Small TNOs, with diameters $\lesssim500$~km, typically do not retain volatile ices on their surfaces \citep{schaller2007,brown2011}, and their reflectance spectra can be well described by a single slope in the optical \citep{fornasier2009} and a single slope in the near-infrared \citep{barucci2011}.
This single optical slope is also seen in the correlation between $g-r$ and $g-i$ colors, \citep[e.g.][]{Sheppard2012,ofek2012,seccull2021}.
Based on the optical and near-infrared surface colors of TNOs, a variety of surface classification schemes have been discussed \citep[e.g.][]{fraser2012,peixinho, dalleore2013, peixinho2015, pike2017z, fraser2021}.
These include two TNO surface classifications \citep{fraser2021}, three surface types \citep[e.g][]{fraser2012, pike2017z}, and significantly more surface classifications \citep{dalleore2013}.
The model of two TNO surface classifications proposed by \citet{fraser2021} based on the Colours of the Outer Solar System Origins Survey (Col-OSSOS) photometry reproduces the range of TNO surface colors in both the optical and near-infrared, does not require additional surface types in the optical and near-infrared, and provides a more accurate division between surface types than optical colors alone. 
Additional complexities in the near-ultra violet may warrant further differentiation in the future, but this two surface classification is sufficient for classification in the optical and near-infrared.

The Col-OSSOS project has acquired near-simultaneous multi-band observations of 102 TNOs.
For all of the TNOs, photometry was acquired at Gemini Observatory in $g$ and $r$ bands, using the Gemini Multi-Object Spectrograph \citep[GMOS,][]{hook2004} and $J$ band using the Near Infrared Imager \citep[NIRI,][]{hodapp2003}.
The data acquisition, reduction, and trailed image photometry methods \citep[TRIPPy]{fraser2016phot} are described in detail in \cite{schwamb2019} and summarized, including improvements since the previous publication, in \citet{fraser2021}.
The photometry used in this work uses the values in  \citet{fraser2021}, which presents the results of the six years of the Col-OSSOS photometry observations, including $g-r$, $r-J$, and some $u-g$ and $r-z$ photometry.
Utilizing this full sample, \citet{fraser2021} determined that the surfaces of TNOs fall into two surface classes.
Because the optical and near infrared (NIR) colors of TNOs broadly follow the reddening line, or curve of constant spectral slope, \citet{fraser2021} recommends a re-projection of the colors into a principal component space, where one component is the line integral distance along the reddening line (PC$^1$), and the second component is the distance away from the reddening line (PC$^2$).
In this projection of $grJ$, the division between the two surface classifications, FaintIR and BrightIR, is a statistically significant linear gap in one of the principle component axes \citep{fraser2021}.
\citet{fraser2021} utilized the Hartigan DIP test of bimodality \citep{hartigan1985}, which for objects with color uncertainties $<0.1$ magnitudes (the approximate gap width) has a $<2$\% chance of being drawn from a unimodal distribution.
See \citet{fraser2021} for a detailed discussion of the significance of the gap and statistical tests used.
\citet{fraser2021} found similar results for other color data using different optical and NIR bands, including additional Col-OSSOS bands and Hubble/WFC3 Test of Surfaces in the Outer Solar System \citep{fraser2012}.
In the $g$, $r$, and $J$ band data, at low values of $PC^1_{grJ}<0.4$, the gap is less apparent and the spectral slopes of these objects more closely resemble the BrightIR surface type even for the few objects with $PC^2_{grJ}<-0.13$ and $PC^1_{grJ}<0.4$.
The FaintIR surface classification has a limited range of optical colors in the red, and includes the vast majority of the objects typically referred to as cold classical TNOs, which have low eccentricity and low inclination orbits.
The BrightIR surface classification is closer to the reddening curve and explores the full range of optical and NIR surfaces.
The range of surface colors in the two classifications can be reproduced using a mixture with one neutral and two different red materials as was done in \citet{fraser2012}, see \citet{fraser2021} for details for the Col-OSSOS sample.
Multiple different compositional materials with different surface reflectivity would be expected if the two surface classifications formed in different regions of the proto-planetesimal disk (this model is agnostic as to the specific material composition and only uses a surface reflectivity).
The two surface classifications of \citet{fraser2021} provides a powerful diagnostic tool for determining the current orbital distribution of the two surface classifications and the implications of this distribution on the formation and evolution of the outer Solar System.

The presence or absence of the FaintIR surface classification in various dynamically excited orbits, particularly in the mean motion resonances of Neptune, can be diagnostic of different modes of planetary migration and different initial disk distributions.
The resonances include objects which are long-term stable as well as objects which temporarily stick into resonance \citep[e.g.][]{lykawkamukai07}.
If Neptune migrated smoothly outward \citep[e.g.][]{malhotra1995}, its resonances would sweep across the cold classical region and deeply trap a significant number of stable FaintIR objects into the mean motion resonances near the cold classical belt.
Some resonances (e.g. the 2:1 presently at 47.7~au) would sweep through portions of the cold disk that are observed today to contain FaintIR objects, while the 3:2 ($a\approx39.4$~au) at the inner edge of today's cold disk would sweep a region that today is dominated by the excited TNO population.
The current-day absence of a cold disk of FaintIR objects interior to the 3:2 could result from the sweeping of sweeping of secular resonances currently located just exterior to the 3:2 (the $\nu_8$ eccentricity secular resonance and the $\nu_{18}$ inclination secular resonance) during migration \citep[see, e.g.][]{volkmalhotra2019}, from the sweeping of the 3:2 itself, or from a primordial absence of such objects; the distribution of surfaces in the 3:2 and other resonances can help probe this question of the primordial extent of the FaintIR surface type.
TNOs which scattered during planetary migration \citep[e.g.][]{Gomes2005,levison2008,Nesvorny2018} or currently stick in the resonances are captured from the bulk dynamically excited population of TNOs, which may include FaintIR and BrightIR surfaces.
The surface classifications of resonant TNOs constrain the capture methods that populated the mean motion resonances of Neptune, and constrains the locations of these source populations.

Recent work has attempted to reproduce the color distribution of TNOs, assuming that the different surface classifications formed in different regions of the proto-planetesimal disk.
These works have typically assumed that the cold classical TNOs have a distinct surface type, different from the dynamically excited TNOs, and often classifies dynamically excited TNO surface types purely based on optical colors (frequently referred to as `red' and `very red').
Utilizing only optical colors replicates the bulk of the BrightIR and FaintIR groups, but results in some contamination, particularly the inclusion of optically very red BrightIR objects in the FaintIR group.
The distribution of red and very red TNO surfaces has been reported to have a correlation between inclination \citep[for a recent analysis, see][]{marsset2019} and eccentricity \citep{alidib2021} to surface color, which is inferred to be related to differences of formation location and the effects of planetary migration on the populations.
The correlation between inclination and surface classification holds when the FaintIR/BrightIR classification system is used \citep{marsset2022}.
\citet{marsset2019} also identified a dependence between the final inclinations of TNOs and their formation distances.
A dependence between color, inclination, and eccentricity has been reproduced in some inward and outward planet migration scenario simulations \citep{pirani2021}.
\citet{nesvorny2020} used an N-body integration of several Neptune migration scenarios \citep{nesvorny_morbidelli2012} with a large disk of test particles, and compared the final orbital and color distributions of test particles for various assumptions about the initial surface density profile of the planetesimal disk and the location in that disk that separates where the red and very red TNOs formed.
\citet{nesvorny2020} compared the red and very red population distributions from these simulations to the then-available observational constraints, and they determined that the observed distribution was consistent with a transition in the initial planetesimal disk from an inner red population to an outer very red population occurring between 30--40 au.
\citet{buchanan2022} tested initial particle distributions of red and very red TNOs distributed within 30 au, and found both combinations (red or very red as the inner type, and the other as the outer type) to provide an acceptable match to the observed current color distribution, with inner/outer color transitions at 27-28 au.
These analyses all utilized optical colors to differentiate between red and very red surfaces instead of a combination of optical and near-infrared slopes, however, a significant fraction of the objects can be properly classified with only the optical slope.
As a result, we expect that with a proper classification into FaintIR and BrightIR surfaces these results can be further refined.

In this work we utilize the FaintIR/BrightIR surface classifications to examine the TNOs trapped in mean motion resonances with Neptune, as the presence of different surface classifications in the resonances provide constraints on the specifics of planetary migration and the initial proto-planetesimal disk.  
We explore the characteristics of the resonant Col-OSSOS TNOs (Section \ref{sec:resonant_sample}), the FaintIR and BrightIR surface classification-distribution within the different resonances (Section \ref{sec:surfaces}), and the implications of this surface-classification distribution for the formation and evolution of the outer Solar System (Section \ref{sec:Discussion}).

\section{Col-OSSOS TNO Sample}
\label{sec:resonant_sample}

The Col-OSSOS sample of $g-r$ and $r-J$ photometry of 102 TNOs includes 37 resonant TNOs in a variety of different resonant orbits.
This work utilizes the same $g-r$ and $r-J$ colors of the Col-OSSOS objects as \citet{fraser2021}.
The photometry and orbital classifications of these targets are listed in Table \ref{tab:targets}, which is sorted by semi-major axis so that the members of each resonance are listed together.
Col-OSSOS was designed to measure a magnitude complete sub-sample of the Outer Solar System Origins Survey (OSSOS) discoveries brighter than $m_r<23.6$ in the OSSOS E, H, L, O, S, and T blocks \citep{bannister2018}.
There is currently one Centaur which meet the Col-OSSOS selection criteria, but does not have Col-OSSOS $g-r$ and $r-J$ colors measured, because the object is currently in the galactic plane.
The object without current photometry is included in Table \ref{tab:targets} with `--' indicated for its Surface Class.
The single unmeasured Centaur should not introduce any significant selection biases that could impact the apparent color distribution of the Col-OSSOS targets, and in particular will not affect our interpretation of the surface distribution of resonant TNOs.

\startlongtable
\begin{deluxetable}{l l c c c c c c c c c c c}
\tabletypesize{\scriptsize}
\tablecaption{\label{tab:targets}
Col-OSSOS Resonant TNOs. 
}
\tablehead{ 
\colhead{MPC}  & \colhead{OSSOS} & \colhead{Dynamical} & \colhead{Discovery}& \colhead{$a$} & \colhead{$e$} & \colhead{$i$} & \colhead{$i_{free}$} & \colhead{(degrees)} & \colhead{$H_{r}$} & \colhead{Surface}  & \colhead{$g-r$} & \colhead{$r-J$}  \\
\colhead{name} & \colhead{name}    & \colhead{Class} & \colhead{mag}  & \colhead{(au)}    & \colhead{}   & \colhead{(degrees)}   & \colhead{\citetalias{huang2022}} & \colhead{\citetalias{vanLaerhoven2019}}  & \colhead{mag}   & \colhead{Class}   & \colhead{}   & \colhead{} \\   
}
\decimals
\startdata
%\startlongtable
%\begin{deluxetable}{l l c c c c c c c c c c c}
%\tabletypesize{\scriptsize}
%\tablecaption{\label{tab:targets}
%Col-OSSOS Resonant TNOs. 
%}
%\tablehead{ 
%\colhead{MPC}  & \colhead{OSSOS} & \colhead{Dynamical} & \colhead{Discovery}& \colhead{$a$} & \colhead{$e$} & \colhead{$i$} & \multicolumn{2}{c}{\colhead{$i_{free}$ (degrees)}} & \colhead{$H_{r}$} & \colhead{Surface}  & \colhead{$g-r$} & \colhead{$r-J$}  \\
%\colhead{name} & \colhead{name}    & \colhead{Class} & \colhead{mag}  & \colhead{(au)}    & \colhead{}   & \colhead{(degrees)}   & \colhead{\citetalias{huang2022}} & \colhead{\citetalias{vanLaerhoven2019}}  & \colhead{mag}   & \colhead{Class}   & \colhead{}   & \colhead{} \\   
%}
%\decimals
%\startdata
2015 RV245 & o5s05 & Centaur & 23.21$\pm$0.040 & 21.98 & 0.48  & 15.39  & --    & -- & 10.10 & BrightIR & 0.670$\pm$0.018 & 1.402$\pm$0.088\\ 
2013 JC64 & o3o01 & Centaur & 23.39$\pm$0.064 & 22.14 & 0.38 & 32.02 & --    & -- & 11.95 & -- & -- & --\\ 
2014 UJ225 & o4h01 & Centaur & 22.74$\pm$0.119 & 23.20  & 0.38 & 21.32 & --    & -- & 10.29 & BrightIR & 0.627$\pm$0.012 & 1.238$\pm$0.105\\ 
2015 RD277 & o5t03 & Centaur & 23.27$\pm$0.040 & 25.97 & 0.29 & 18.85 & -- & -- & 10.48 & FaintIR & 0.906$\pm$0.033 & 1.722$\pm$0.134\\ 
2015 RK277 & o5s01 & Jupiter Coupled & 23.36$\pm$0.070 & 26.91 & 0.80 & 9.53 & --    & -- & 15.29 & BrightIR & 0.527$\pm$0.015 & 1.355$\pm$0.052\\ 
2015 RU245 & o5t04 & scattering & 22.99$\pm$0.037 & 30.99  & 0.29 & 13.75 & --    & -- & 9.32 & FaintIR & 0.884$\pm$0.018 & 1.572$\pm$0.066\\ 
2002 GG166 & o3e01 & scattering & 21.50$\pm$0.087 & 34.42 & 0.59 & 7.71 & --    & -- & 7.73 & BrightIR & 0.591$\pm$0.013 & 1.496$\pm$0.050\\ 
2014 UX228 & o4h18 & 4:3 resonant & 23.11$\pm$0.054 & 36.35  & 0.17 & 20.66 & --    & -- & 7.35 & BrightIR & 0.497$\pm$0.022 & 1.488$\pm$0.060\\ 
2013 US15 & o3l09 & 4:3 resonant & 23.24$\pm$0.156 & 36.38  & 0.07 & 2.02 & --    & -- & 7.78 & FaintIR & 1.050$\pm$0.015 & 1.491$\pm$0.065\\ 
2014 UD229 & o4h13 & 4:3 resonant & 23.54$\pm$0.074 & 36.39 & 0.15 & 6.85 & --    & -- & 8.18 & BrightIR & 0.691$\pm$0.016 & 1.304$\pm$0.076\\ 
2006 QP180 & o4h67PD & scattering & 23.07$\pm$0.044 & 38.08 & 0.65 & 4.96 & --    & -- & 9.49 & BrightIR & 0.950$\pm$0.082 & 2.069$\pm$0.099\\ 
2013 SZ99 & o3l15 & hot classical & 23.54$\pm$0.127 & 38.28 & 0.02 & 19.84 & --    & -- & 7.65 & BrightIR & 0.592$\pm$0.019 & 1.538$\pm$0.083\\ 
2014 UH225 & o4h29 & hot classical & 23.31$\pm$0.064 & 38.64 & 0.04 & 29.53 & --    & -- & 7.30 & BrightIR & 0.532$\pm$0.017 & 1.628$\pm$0.056\\ 
2015 RU277 & o5s07 & 3:2 resonant & 23.52$\pm$0.040 & 39.33 & 0.25 & 16.27 & -- & -- & 8.69 & BrightIR & 0.603$\pm$0.018 & 1.243$\pm$0.066\\ 
2001 RX143 & o4h76PD & 3:2 resonant & 22.84$\pm$0.062 & 39.34 & 0.30 & 19.25 & --    & -- & 6.42 & FaintIR & 0.843$\pm$0.038 & 1.315$\pm$0.109\\ 
2007 JF43 & o3o20PD & 3:2 resonant & 21.15$\pm$0.020 & 39.35 & 0.18 & 15.08 & --  & -- & 5.27 & BrightIR & 0.978$\pm$0.011 & 1.953$\pm$0.048\\ 
2013 JD65 & o3o15 & 3:2 resonant & 23.48$\pm$0.073 & 39.37 & 0.09 & 13.02 & -- & -- & 7.90 & BrightIR & 0.790$\pm$0.030 & 1.881$\pm$0.087\\ 
2013 JJ65 & o3o27 & 3:2 resonant & 23.40$\pm$0.076 & 39.37  & 0.26 & 19.82 & -- & -- & 7.22 & FaintIR & 1.076$\pm$0.027 & 1.695$\pm$0.075\\ 
2013 JB65 & o3o09 & 3:2 resonant & 23.23$\pm$0.060 & 39.40 & 0.19 & 24.90 & -- & -- & 8.13 & BrightIR & 0.725$\pm$0.018 & 1.101$\pm$0.090\\ 
2003 SR317 & o3l13PD & 3:2 resonant & 23.36$\pm$0.084 & 39.43 & 0.17 & 8.35 & -- & -- & 7.66 & BrightIR & 0.646$\pm$0.012 & 1.359$\pm$0.065\\ 
2013 GH137 & o3e02 & 3:2 resonant & 23.34$\pm$0.140 & 39.44 & 0.23 & 13.47 & -- &   -- & 8.32 & BrightIR & 0.713$\pm$0.032 & 1.757$\pm$0.097\\ 
2014 UO229 & o4h11 & 3:2 resonant & 23.55$\pm$0.070 & 39.45 & 0.16 & 10.09 & -- &   -- & 8.25 & BrightIR & 0.728$\pm$0.022 & 1.162$\pm$0.078\\ 
2014 UV228 & o4h09 & 3:2 resonant & 23.48$\pm$0.082 & 39.49 & 0.23 & 10.13 & -- &   -- & 8.49 & BrightIR & 0.592$\pm$0.021 & 1.461$\pm$0.060\\ 
2013 GJ137 & o3e04 & 3:2 resonant & 23.39$\pm$0.163 & 39.50 & 0.27 & 16.87 & -- & -- & 8.25 & BrightIR & 0.611$\pm$0.019 & 1.675$\pm$0.065\\ 
2014 UF228 & o4h70 & 3:2 resonant & 22.70$\pm$0.044 & 39.55  & 0.22 & 12.60 & -- & -- & 7.77 & BrightIR & 0.611$\pm$0.024 & 1.380$\pm$0.072\\ 
2014 UX229 & o4h05 & 3:2 resonant & 22.25$\pm$0.038 & 39.63 & 0.34 & 15.97 & -- & -- & 8.04 & BrightIR & 0.653$\pm$0.013 & 1.463$\pm$0.090\\ 
1995 QY9 & o4h69PD & 3:2 resonant & 22.38$\pm$0.062 & 39.64 & 0.26 & 4.83 & -- & -- & 7.68 & BrightIR & 0.737$\pm$0.021 & 1.464$\pm$0.057\\ 
2010 TJ182 & o4h07 & 3:2 resonant & 22.28$\pm$0.020 & 39.65 & 0.28 & 9.50 & -- & -- & 7.68 & BrightIR & 0.559$\pm$0.018 & 1.340$\pm$0.064\\ 
2013 JN65 & o3o28 & hot classical & 23.42$\pm$0.213 & 40.67 & 0.01 & 19.64 & 17.9271 & 18.290 & 7.23 & BrightIR & 0.578$\pm$0.013 & 1.634$\pm$0.062\\ 
2013 GO137 & o3e29 & hot classical & 23.46$\pm$0.080 & 41.42  & 0.09 & 29.25 & 28.5871 & 26.836 & 7.09 & BrightIR & 0.768$\pm$0.027 & 1.727$\pm$0.059\\ 
2001 RY143 & o4h48 & hot classical & 23.54$\pm$0.080 & 42.08 & 0.16 & 6.91 & 8.5056 & 6.262 & 6.80 & BrightIR & 0.892$\pm$0.027 & 1.890$\pm$0.074\\ 
2001 QF331 & o3l06PD & 5:3 resonant & 22.71$\pm$0.067 & 42.25 & 0.25 & 2.67 & -- &-- & 7.56 & FaintIR & 0.830$\pm$0.025 & 1.579$\pm$0.074\\ 
2013 UM17** & o3l29PD & hot classical & 23.56$\pm$0.094 & 42.48 & 0.08 & 12.99 & 7.29 & 13.968 & 7.29 & BrightIR & 0.751$\pm$0.048 & 1.246$\pm$0.179\\ 
2013 UQ15 & o3l77 & hot classical & 22.93$\pm$0.124 & 42.77  & 0.11 & 27.34 & 27.3663 & 28.183 & 6.07 & BrightIR & 0.473$\pm$0.032 & 0.942$\pm$0.120\\ 
2013 GW137 & o3e54 & cold classical & 23.61$\pm$0.109 & 42.86 & 0.06 & 5.02 & 3.4976&3.467 & 7.50 & FaintIR & 0.946$\pm$0.060 &   1.892$\pm$0.134\\ 
2013 GC138 & o3e32 & cold classical & 23.52$\pm$0.144 & 42.89 & 0.05 & 3.02 & 1.0871 & 1.069 & 7.02 & BrightIR & 1.021$\pm$0.056 & 1.027$\pm$0.253\\ 
2015 RG277 & o5s45 & hot classical & 23.15$\pm$0.030 & 42.96 & 0.01 & 12.09 & 12.6632 &13.068 & 6.79 & FaintIR & 0.950$\pm$0.018 & 1.642$\pm$0.061\\ 
2013 GB138 & o3e38 & cold classical & 23.60$\pm$0.107 & 42.98 & 0.05 & 2.79 & 3.7896 & 3.772 & 7.01 & FaintIR & 0.921$\pm$0.049 & 1.758$\pm$0.126\\ 
2001 FK185 & o3e20PD & cold classical & 23.09$\pm$0.215 & 43.24 & 0.04 & 1.17& 0.9268 & 0.966 & 6.82 & BrightIR & 0.833$\pm$0.033 & 1.773$\pm$0.077\\ 
2015 RB281 & o5s36 & cold classical & 23.58$\pm$0.068 & 43.26 & 0.05 & 2.21 & 4.0507 &4.055 & 7.36 & FaintIR & 0.885$\pm$0.027 & 1.498$\pm$0.07\\ 
2013 GX137 & o3e28 & cold classical & 23.17$\pm$0.096 & 43.29 & 0.06 & 4.13 & 2.6515&2.677 & 6.82 & FaintIR & 0.983$\pm$0.028 & 1.460$\pm$0.081\\ 
2013 UO15 & o3l50 & cold classical & 23.20$\pm$0.064 & 43.33  & 0.05 & 3.73 &2.0387 &2.017 & 6.69 & FaintIR & 0.955$\pm$0.019 & 1.705$\pm$0.055\\ 
2014 UD225 & o4h45 & cold classical & 23.09$\pm$0.050 & 43.36 & 0.13 & 3.66 & 4.0439 &4.364 & 6.63 & BrightIR & 0.712$\pm$0.016 & 1.251$\pm$0.087\\ 
2014 UK225 & o4h19 & hot classical & 23.23$\pm$0.061 & 43.52 & 0.13 & 10.69 & 12.2369&12.330 & 7.43 & FaintIR & 0.978$\pm$0.017 & 1.677$\pm$0.055\\ 
2013 UX18 & o3l69 & cold classical & 23.42$\pm$0.105 & 43.60  & 0.06 & 2.89 &1.061 &1.050 & 6.74 & FaintIR & 0.888$\pm$0.010 & 1.655$\pm$0.093\\ 
2013 GR136 & o3e19 & 7:4 resonant & 23.40$\pm$0.098 & 43.65 & 0.08 & 1.65 & --&-- & 7.20 & BrightIR & 0.717$\pm$0.026 & 1.466$\pm$0.102\\ 
2016 BP81 & o3l39 & 7:4 resonant* & 22.92$\pm$0.051 & 43.68 & 0.08 & 4.18 & -- &-- & 6.55 & BrightIR & 0.572$\pm$0.026 & 1.572$\pm$0.086\\ 
2001 QE298 & o5t11PD & 7:4 resonant & 23.17$\pm$0.040 & 43.71 & 0.16 & 3.66 &-- &-- & 7.38 & FaintIR & 0.834$\pm$0.018 & 1.494$\pm$0.063\\ 
2013 GP137 & o3e35 & cold classical & 23.48$\pm$0.133 & 43.71 & 0.03 & 1.75 & 1.6862&1.700 & 6.94 & FaintIR & 0.943$\pm$0.033 & 1.268$\pm$0.100\\ 
2014 UE225 & o4h50 & hot classical & 22.67$\pm$0.037 & 43.71 & 0.07 & 4.49 & 6.189&6.207 & 5.99 & FaintIR & 1.040$\pm$0.017 & 1.817$\pm$0.067\\ 
2013 SP99 & o3l32 & cold classical & 23.47$\pm$0.077 & 43.78 & 0.06 & 0.79 & 1.0571 &1.058 & 7.23 & FaintIR & 0.977$\pm$0.020 & 1.606$\pm$0.067\\ 
2013 GV137 & o3e43 & cold classical & 23.42$\pm$0.284 & 43.79 & 0.08 & 3.20 &1.428 &1.528 & 6.67 & FaintIR & 0.922$\pm$0.022 & 1.275$\pm$0.124\\ 
2015 RO281 & o5s21 & cold classical & 23.17$\pm$0.054 & 43.85 & 0.17 & 2.37 & 5.199 & 4.139 & 7.34 & FaintIR & 0.965$\pm$0.034 & 1.396$\pm$0.116\\ 
2013 GS137 & o3e16 & cold classical & 23.47$\pm$0.142 & 43.87  & 0.10 & 2.60 & 0.9472&0.881 & 7.44 & FaintIR & 1.010$\pm$0.022 & 1.720$\pm$0.080\\ 
2004 HJ79 & o3e37PD & cold classical & 23.37$\pm$0.087 & 43.96 & 0.05 & 3.32 & 2.284 &2.288 & 6.81 & FaintIR & 0.953$\pm$0.020 & 1.595$\pm$0.068\\ 
2013 GF138 & o3e34PD & cold classical & 23.57$\pm$0.102 & 44.04 & 0.02 & 0.55 & 1.8315 &1.855 & 7.05 & FaintIR & 1.073$\pm$0.026 & 1.708$\pm$0.065\\ 
2013 GN137 & o3e22 & cold classical & 22.97$\pm$0.093 & 44.10 & 0.07 & 2.76 & 1.0996&1.134 & 6.70 & FaintIR & 1.053$\pm$0.010 & 1.742$\pm$0.070\\ 
2013 GM137 & o3e51 & hot classical & 23.32$\pm$0.227 & 44.10 & 0.08 & 22.46 & 21.8853&21.477 & 6.90 & BrightIR & 0.597$\pm$0.041 & 1.194$\pm$0.129\\ 
2004 EU95 & o3e27PD & cold classical & 23.10$\pm$0.099 & 44.15 & 0.04 & 2.82 & 1.3554 &1.347 & 6.77 & FaintIR & 0.969$\pm$0.023 & 1.813$\pm$0.064\\ 
2013 SQ99 & o3l76 & cold classical & 23.10$\pm$0.061 & 44.15 & 0.09 & 3.47 &2.21 &2.266 & 6.35 & FaintIR & 0.971$\pm$0.023 & 1.696$\pm$0.072\\ 
2015 RT245 & o5t31 & cold classical & 22.87$\pm$0.043 & 44.39  & 0.08 & 0.96 & 2.3451&2.352 & 6.57 & FaintIR & 0.944$\pm$0.047 & 1.585$\pm$0.122\\ 
2014 UM225 & o4h31 & 9:5 resonant & 23.25$\pm$0.064 & 44.48 & 0.10 & 18.30 & -- &-- & 7.21 & FaintIR & 0.792$\pm$0.014 & 1.529$\pm$0.061\\ 
2013 GT137 & o3e31 & cold classical & 23.55$\pm$0.134 & 44.59  & 0.11 & 2.29 & 2.0614 &2.186 & 7.10 & FaintIR & 1.040$\pm$0.038 & 1.788$\pm$0.086\\ 
2006 QF181 & o3l60 & cold classical & 23.29$\pm$0.074 & 44.81 & 0.08 & 2.66  &1.9568& 1.939 & 6.79 & FaintIR & 0.898$\pm$0.025 & 1.539$\pm$0.072\\ 
2013 GU137 & o3e25 & cold classical & 23.59$\pm$0.087 & 44.84 & 0.07 & 4.97 & 4.2313 & 4.219 & 7.30 & FaintIR & 0.987$\pm$0.072 & 1.648$\pm$0.152\\ 
2013 GY137 & o3e53 & cold classical & 23.50$\pm$0.232 & 44.89 & 0.10 & 5.31 &4.4003 & 4.428 & 7.29 & BrightIR & 0.879$\pm$0.027 & 1.852$\pm$0.150\\ 
2013 UM15 & o3l57 & 11:6 resonant & 23.37$\pm$0.084 & 45.04  & 0.07  & 1.84  & -- &-- & 6.81 & FaintIR & 1.048$\pm$0.010 & 1.729$\pm$0.068\\ 
2013 UN15 & o3l63 & cold classical & 23.62$\pm$0.180 & 45.13  & 0.05  & 3.36  & 3.3263&3.337 & 7.00 & FaintIR & 1.068$\pm$0.023 & 1.6$\pm$0.062\\ 
2013 EM149 & o3e30PD & cold classical & 22.99$\pm$0.054 & 45.26  & 0.06  & 2.63  &2.784& 2.795 & 6.59 & FaintIR & 0.958$\pm$0.021 & 1.651$\pm$0.060\\ 
2013 GQ137 & o3e21 & cold classical & 23.40$\pm$0.093 & 45.69  & 0.13  & 2.85  & 1.1202&1.198 & 7.12 & BrightIR & 0.892$\pm$0.022 & 1.874$\pm$0.062\\ 
2013 HR156 & o3e49 & 15:8 resonant & 23.54$\pm$0.090 & 45.72  & 0.19  & 20.41  & -- &-- & 7.72 & BrightIR & 0.586$\pm$0.027 & 1.363$\pm$0.112\\ 
2013 UL15 & o3l43 & cold classical & 23.05$\pm$0.109 & 45.79  & 0.10  & 2.02  & 0.4755 &0.532 & 6.62 & FaintIR & 0.895$\pm$0.031 & 1.482$\pm$0.076\\ 
2014 UN228 & o4h75 & hot classical & 23.37$\pm$0.113 & 45.87  & 0.17  & 24.02  & 23.8737 &23.662 & 7.46 & BrightIR & 0.601$\pm$0.025 & 1.459$\pm$0.085\\ 
2003 SP317 & o5t34PD & 17:9 resonant* & 23.48$\pm$0.078 & 45.96 & 0.17 & 5.08 & -- & -- & 7.06 & BrightIR & 0.958$\pm$0.047 & 1.188$\pm$0.153\\ 
2010 RE188 & o3l18 & hot classical & 22.27$\pm$0.051 & 46.01  & 0.15  & 6.75  & 7.1516 &7.079 & 6.19 & BrightIR & 0.677$\pm$0.015 & 1.429$\pm$0.079\\ 
2013 JR65 & o3o21 & hot classical & 23.51$\pm$0.123 & 46.20  & 0.19  & 11.71  & 10.048 &10.079 & 7.53 & BrightIR & 0.450$\pm$0.010 & 1.593$\pm$0.073\\ 
2013 SA100 & o3l79 & hot classical & 22.81$\pm$0.044 & 46.30  & 0.17  & 8.48  & 7.8874 & 8.001 & 5.77 & BrightIR & 0.649$\pm$0.014 & 1.542$\pm$0.057\\ 
2014 UL225 & o4h20 & hot classical & 23.03$\pm$0.069 & 46.34  & 0.20  & 7.95  & 9.2175 &9.075 & 7.24 & BrightIR & 0.556$\pm$0.030 & 0.769$\pm$0.128\\ 
2013 JX67 & o3o51 & hot classical & 22.75$\pm$0.037 & 46.39  & 0.13  & 10.50  & 8.8721 &8.897 & 6.49 & BrightIR & 0.697$\pm$0.016 & 1.399$\pm$0.062\\ 
2001 FO185 & o3e23PD & hot classical & 23.37$\pm$0.077 & 46.45  & 0.12  & 10.64  & 11.2669 &11.289 & 7.09 & BrightIR & 0.861$\pm$0.022 & 1.866$\pm$0.068\\ 
2013 UP15 & o3l46 & cold classical & 23.72$\pm$0.097 & 46.61  & 0.08  & 2.47  & 0.9583 &0.990 & 7.26 & BrightIR & 0.884$\pm$0.020 & 1.902$\pm$0.087\\ 
2015 RJ277 & o5s32 & hot classical & 23.21$\pm$0.040 & 46.70 & 0.20  & 5.52  & 5.9227 &6.175 & 7.12 & BrightIR & 0.638$\pm$0.020 & 1.175$\pm$0.108\\ 
2013 GG138 & o3e44 & hot classical & 23.26$\pm$0.093 & 47.46  & 0.03  & 24.61  & 25.2462 &25.348 & 6.34 & FaintIR & 1.090$\pm$0.031 & 1.854$\pm$0.075\\ 
2013 GW136 & o3e05 & 2:1 resonant & 22.69$\pm$0.066 & 47.74  & 0.34  & 6.66  & --&-- & 7.42 & BrightIR & 0.724$\pm$0.019 & 1.697$\pm$0.065\\ 
2013 JE64 & o3o18 & 2:1 resonant & 23.56$\pm$0.150 & 47.79  & 0.29  & 8.34  &-- &-- & 7.94 & FaintIR & 0.954$\pm$0.120 & 1.472$\pm$0.247\\ 
2013 GX136 & o3e55 & 2:1 resonant & 23.41$\pm$0.134 & 48.00  & 0.25  & 1.10  & --&-- & 7.67 & BrightIR & 0.734$\pm$0.023 & 1.645$\pm$0.071\\ 
2013 GQ136 & o3e45 & hot classical & 23.59$\pm$0.102 & 48.73 & 0.17 & 2.03 & -- & -- & 6.13 & FaintIR & 1.029$\pm$0.051 & 1.350$\pm$0.205\\ 
2014 UQ229 & o4h03 & scattering & 22.69$\pm$0.213 & 49.90 & 0.78  & 5.68  & --&-- & 9.55 & BrightIR & 0.936$\pm$0.018 & 1.996$\pm$0.062\\ 
2015 RU278 & o5s52 & 11:5 resonant* & 23.26$\pm$0.040 & 50.85 & 0.25 & 27.20 & -- & -- & 6.55 & BrightIR & 0.612$\pm$0.029 & 1.537$\pm$0.118\\ 
2013 JK64 & o3o11 & 5:2 resonant & 22.94$\pm$0.045 & 55.25  & 0.41  & 11.08  & --&-- & 7.69 & BrightIR & 0.978$\pm$0.031 & 1.889$\pm$0.059\\ 
2014 US229 & o4h14 & 5:2 resonant & 23.18$\pm$0.083 & 55.26  & 0.40  & 3.90  & --&-- & 7.95 & BrightIR & 0.628$\pm$0.020 & 1.418$\pm$0.068\\ 
2013 GY136 & o3e09 & 5:2 resonant & 22.94$\pm$0.051 & 55.54 & 0.41  & 10.88  & --&-- & 7.32 & BrightIR & 0.513$\pm$0.018 & 1.191$\pm$0.109\\ 
2013 UR15 & o3l01 & scattering & 23.06$\pm$0.065 & 55.82 & 0.72  & 22.25  & --&-- & 10.89 & BrightIR & 0.667$\pm$0.023 & 1.641$\pm$0.093\\ 
2015 RW245 & o5s06 & scattering & 22.90$\pm$0.030 & 56.48 & 0.53  & 13.30  & --&-- & 8.53 & BrightIR & 0.683$\pm$0.020 & 1.543$\pm$0.075\\ 
2013 JL64 & o3o29 & detached & 23.26$\pm$0.123 & 56.77  & 0.37  & 27.67  & --&-- & 7.03 & BrightIR & 0.681$\pm$0.025 & 1.428$\pm$0.132\\ 
2013 JH64 & o3o34 & 11:4 resonant* & 22.70$\pm$0.037 & 59.20 & 0.38  & 13.73  & --&-- & 5.60 & BrightIR & 0.697$\pm$0.019 & 1.470$\pm$0.079\\ 
2014 UA225 & o5t09PD & detached & 22.50$\pm$0.020 & 67.76  & 0.46  & 3.58  & --&-- & 6.74 & FaintIR & 0.950$\pm$0.018 & 1.464$\pm$0.060\\ 
2015 RR245 & o5s68 & 9:2 resonant* & 21.76$\pm$0.013 & 81.7 & 0.58  & 7.55  & --&-- & 3.60 & BrightIR & 0.765$\pm$0.012 & 1.188$\pm$0.083\\ 
2013 GZ136 & o3e11 & scattering & 23.60$\pm$0.098 & 86.74 & 0.61  & 18.36  & --&-- & 7.86 & BrightIR & 0.714$\pm$0.020 & 1.608$\pm$0.122\\ 
2004 PB112 & o5s16PD & 27:4 resonant* & 22.99$\pm$0.027 & 107.52 & 0.67  & 15.43  &-- &-- & 7.39 & FaintIR & 0.824$\pm$0.022 & 1.422$\pm$0.059\\ 
2007 TC434 & o4h39 & 9:1 resonant & 23.21$\pm$0.054 & 129.94 & 0.70  & 26.47  &-- &-- & 7.13 & BrightIR & 0.670$\pm$0.015 & 1.499$\pm$0.063\\ 
2013 JO64 & o3o14 & scattering & 23.54$\pm$0.080 & 143.30 & 0.75  & 8.58  &-- &-- & 8.00 & BrightIR & 0.548$\pm$0.027 & 1.152$\pm$0.092\\ 
2013 GP136 & o3e39 & detached & 23.07$\pm$0.067 & 150.2 & 0.73  & 33.54  &-- &-- & 6.42 & BrightIR & 0.769$\pm$0.020 & 1.633$\pm$0.066\\ 
\enddata
\vspace{0.5cm}
Note. The Minor Planet Center (MPC) identification and OSSOS internal name are both provided for all targets.  The Discovery magnitude is the magnitude calculated by the OSSOS project for the discovery images, and solar system absolute magnitude $H_r$ was calculated from the discovery magnitude, and uncertainty is dependent on the uncertainty in discovery magnitude.  The orbital fit parameters (semi-major axis $a$, eccentricity $e$, and inclination $i$) and the object Dynamical Classification are from the OSSOS orbital fits \citep{bannister2018}, and all digits printed are significant.  Insecure resonance classification, where one or both of the variant orbits do not show the resonant behavior seen in the nominal clone, is indicated by a `*'.  The free inclination $i_{free}$ is calculated by \citet[][H2022]{huang2022} and \citet[][VL2019]{vanLaerhoven2019}; the free inclination is used to separate classical TNOs into `cold' and `hot' classes  according to \cite{vanLaerhoven2019}'s results (see main text).  The Surface Class column indicates whether the object is BrightIR or FaintIR based on their PC values (FaintIR: PC$^2_{grJ}>-0.13$ and $PC^1_{grJ}>0.4$).  The $g-r$ and $r-J$ colors are in the Sloan Digital Sky Survey (SDSS) filters \citep{fraser2021}.
The object indicated with `**' was not on the initial target selection list, but with an improved re-measurement of the OSSOS discovery images, this object meets the Col-OSSOS selection criteria.
\end{deluxetable}

\begin{figure}[h!]
    \includegraphics[width=1.\textwidth]{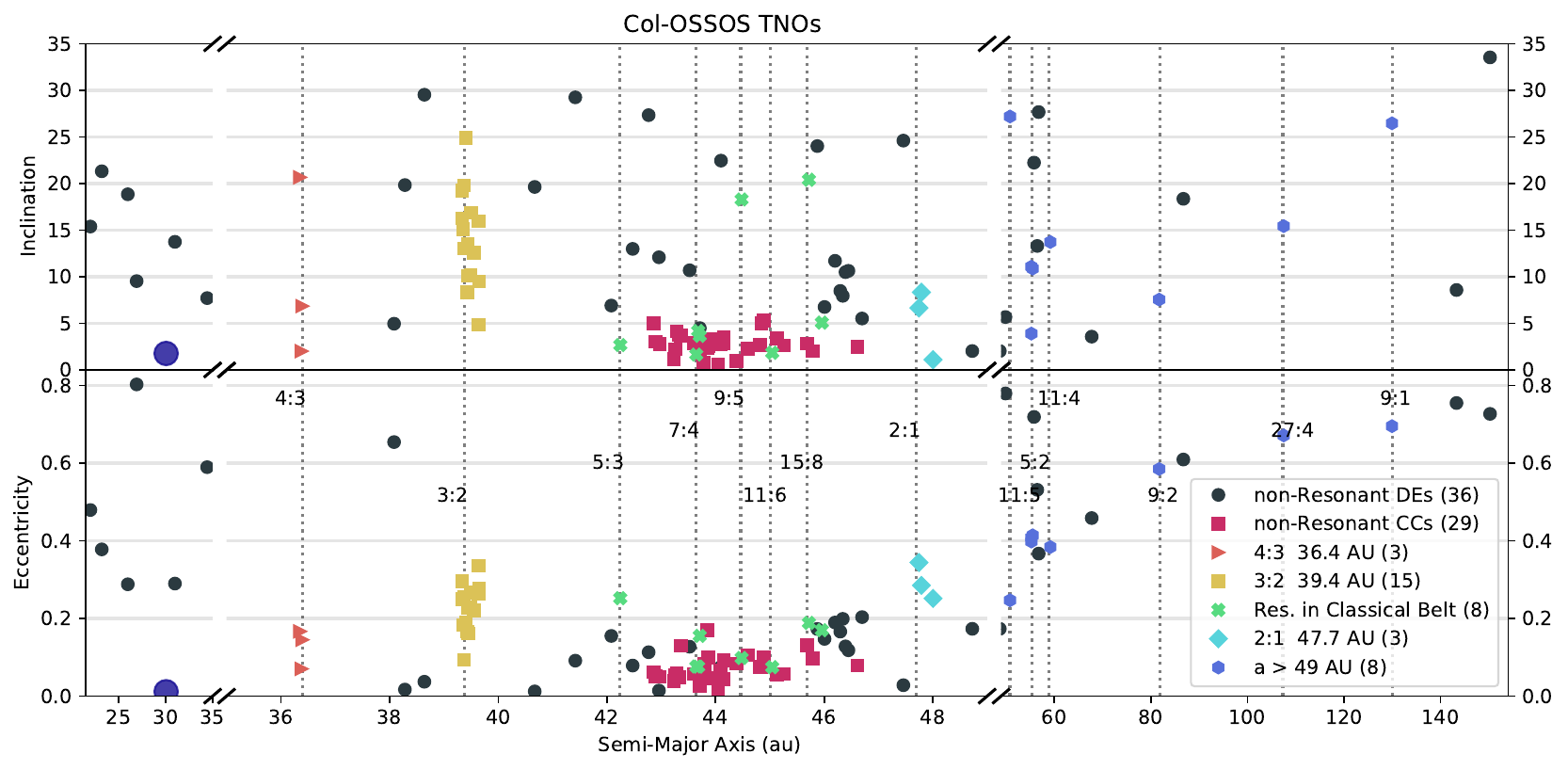}
    \caption{The orbital distribution of Col-OSSOS targets in $a$, $e$, and oscillating $i$ as reported by \citep{bannister2018}.  The semi-major axis is shown with three different scales, as indicated in the x-axis and separated by axis gaps, in order to show the details in the cold classical region as well as the inner and distant objects in the Col-OSSOS sample.  The different dynamical classifications are non-resonant Dynamically Excited objects (DE, black circles), cold classical objects (pink squares), 4:3 resonators (orange triangles), 3:2 resonators (yellow squares), resonant objects within the classical region (green wide `x's, includes all resonant objects within the main classical belt), 2:1 resonators (cyan diamonds), and resonant objects with semi-major axes $a>49$~au (blue hexagons).  The number of objects in each classification is indicated in parentheses in the legend.  All populated resonances are labeled with gray dashed vertical lines.  Neptune is indicated with a large blue circle.
    } \label{fig:aei}
\end{figure}

The TNOs were classified into different dynamical groups utilizing the \cite{gladman2008} scheme as described in the OSSOS data release publications \citep{bannister2016,bannister2018}.
The resonant TNO classifications are determined by numerically integrating the OSSOS orbits (including two clones representing the maximal orbital uncertainty) of each TNO and examining the time histories for libration of any resonant angles \citep[see, e.g.][]{volk2016,Volk2018}.
We use these resonant classifications exactly as identified by OSSOS.
Insecure classifications (where the classification of one or both of the extremal semi-major axis clones does not agree with the nominal) are indicated by a `*' in Table \ref{tab:targets}.
We consider the samples of 4:3, 3:2, and 2:1 resonant TNOs individually.
Because of the small number of Col-OSSOS targets in some of Neptune's resonances, in this work we combine all of the resonant TNOs with semi-major axes between the 3:2 and 2:1 resonances into a single group, which we refer to as ``resonant within the classical belt'', and we combine all of the resonant TNOs beyond the 2:1 resonance (at 47 au) into a group which we refer to as ``resonant $a>49$~au.''
For the remainder of the Col-OSSOS sample, we divide the objects into two groups based on their orbital dynamics: ``cold classical'' and ``dynamically excited'' objects.
The number of TNOs in each category is given in the legend of Figure \ref{fig:aei}.
We utilize a classification scheme for cold classical TNOs and dynamically excited objects based on \citet{vanLaerhoven2019}'s analysis of the inclination distributions of the TNO populations and similar work by \citet{huang2022}.
\citet{vanLaerhoven2019} computed the free inclination (i.e., the orbit's inclination relative to a locally dynamically meaningful reference plane) for classical OSSOS TNOs, and found that the free inclination $i_{free}$ is a much more robust criteria for evaluating the degree of dynamical stirring which the object has experienced (see, e.g., \citealt{Gladman2021} for a review of both the concept of free inclinations and the motivation for using $i_{free}$ to separate cold and excited TNOs).
As a result, $i_{free}$ is a much more effective criteria for classifying objects as dynamically cold than the current osculating ecliptic orbital inclination.
The criteria used to classify cold classical TNOs, or those objects which are most likely to have formed in their current location, are semi-major axes $42.5<a<45$~au and $i_{free}<4^{\circ}$ or $45<a<47$~au with $i_{free}<6^{\circ}$, as the more distant portion of the classical belt appears to have experienced slightly more stirring.
We use the $i_{free}$ values from \citet{vanLaerhoven2019}, given in Table \ref{tab:targets}, to identify cold classical TNOs, and all of the remaining Col-OSSOS objects are grouped into the category of ``non-Resonant Dynamically Excited'' TNOs.
These dynamically excited (DE) TNOs include all objects described in the Table \ref{tab:targets} as Centaurs, hot classical, detached, scattering, and Jupiter coupled.
Figure \ref{fig:aei} shows the orbital distribution of the resonant and non-resonant Col-OSSOS TNOs that have $g-r$ and $r-J$ colors.
The significant number of Plutinos is immediately apparent, as is the motivation for combining the resonant objects within the classical belt into one group and the resonant TNOs beyond the 2:1 resonance into another.
We combine these dynamical classifications with the FaintIR and BrightIR surface classification from \citet{fraser2021} to understand the surface-classification distributions within Neptune's resonances.

\section{The Surfaces of Resonant TNOs}
\label{sec:surfaces}

\begin{figure}[h!]
    \includegraphics[width=.5\textwidth]{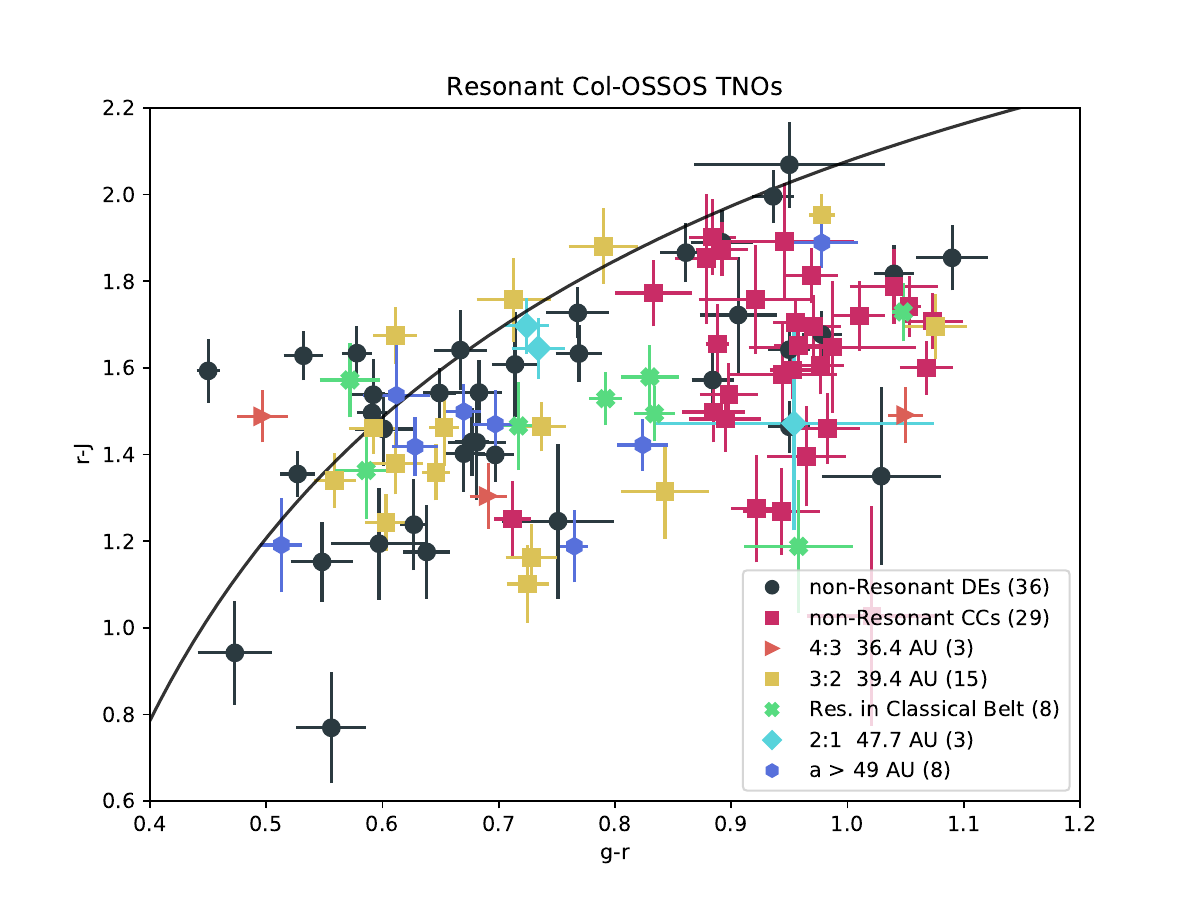}
        \includegraphics[width=.5\textwidth]{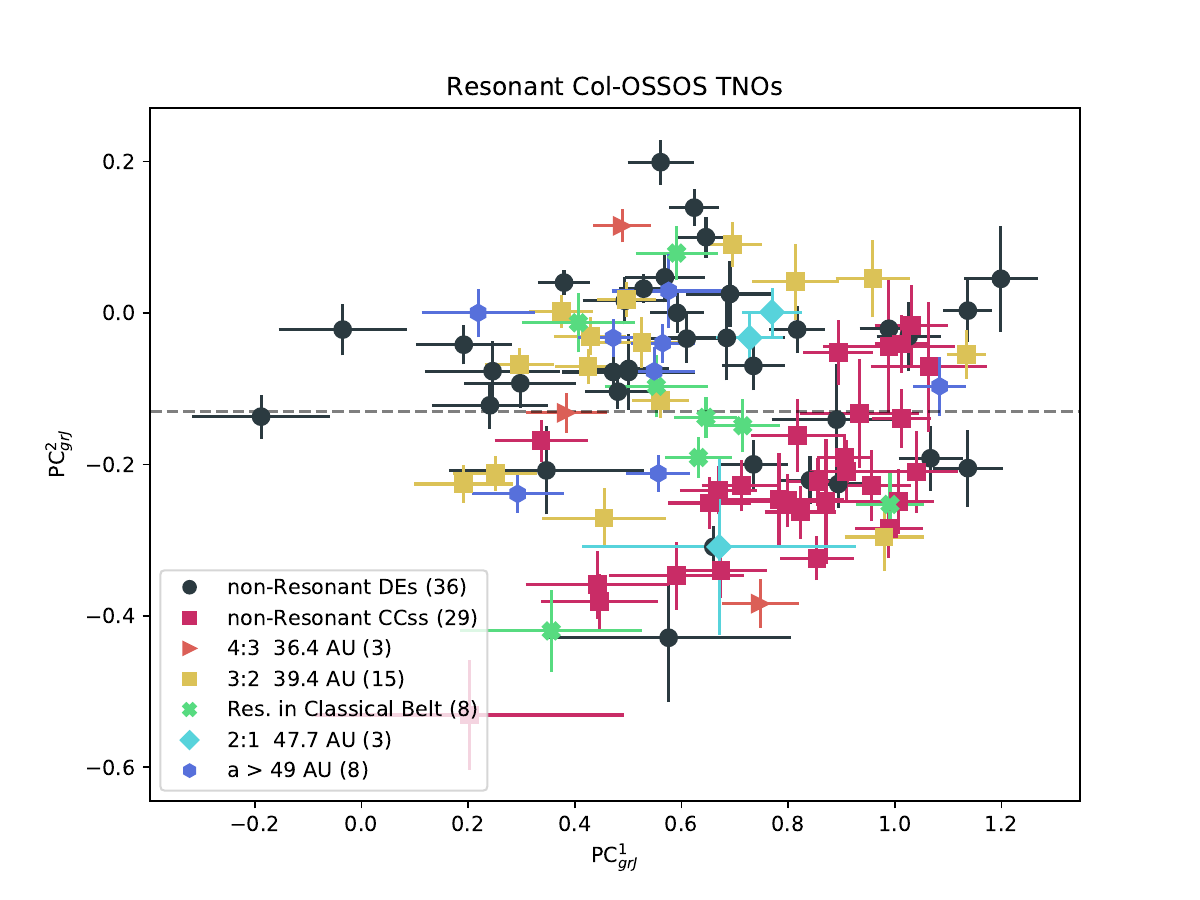}
    \caption{\textbf{Left:} The $r-J$ vs. $g-r$ color distribution of Col-OSSOS TNOs. The black arc is the reddening line, or line of constant spectral slope. The same symbol colors are used here as in Figure \ref{fig:aei}: non-resonant Dynamically excited objects (black circles), cold classical objects (pink squares), 4:3 resonators (orange triangles), 3:2 resonators (yellow squares), resonant objects within the classical region (green wide `x's), 2:1 resonators (cyan diamonds), and resonant objects with semi-major axes $a>49$~au (blue hexagons).  The Plutinos are the largest resonant sample, and span the full range of surface colors seen in the sample. The objects below the gap feature (parallel to the reddening line) are primarily cold classical and resonant TNOs. \textbf{Right:} The color-color plot re-projected into the PC$^1$ and PC$^2$ components, the distance along the reddening line and from the reddening line respectively.  In this re-projection, the gap feature is indicated by the dashed line at PC$^2_{grJ}=-0.13$.  The FaintIR surface classification below this line is dominated by cold classical objects and resonant TNOs.  It includes one 4:3, one 2:1, and several 3:2 resonators, as well as a few dynamically excited objects and a distant $a>49$~au resonator.
    } \label{fig:grJ}
\end{figure}

In the projection color space, the Col-OSSOS sample of TNOs shows only a single bifurcation and gap, into the two surface types BrightIR and FaintIR \citep{fraser2021}.
For the resonant sub-sample discussed in this work, we utilize the same re-projection into the principal components along and away from the reddening line \citep[PC$^1$ and PC$^2$ respectively,][]{fraser2021} to classify the objects into FaintIR and BrightIR surface classifications.
We find that the colors of resonant TNOs span the full range of color-color space in $g-r$ and $r-J$ and in the re-projection.
As we are utilizing the $g-r$ and $r-J$ colors, we refer to the re-projected values as PC$^1_{grJ}$ and PC$^2_{grJ}$.
In the re-projection, the FaintIR and BrightIR are divided primarily based on PC$^2_{grJ}=-0.13$, the dashed line in the right panel of Figure \ref{fig:grJ}, with the small sub-sample which are PC$^2_{grJ}=-0.13$ classified as BrightIR, as these are beyond the statistically significant gap and their spectral slopes more closely resemble that group.

\begin{figure}[h!]
    \includegraphics[width=.9\textwidth]{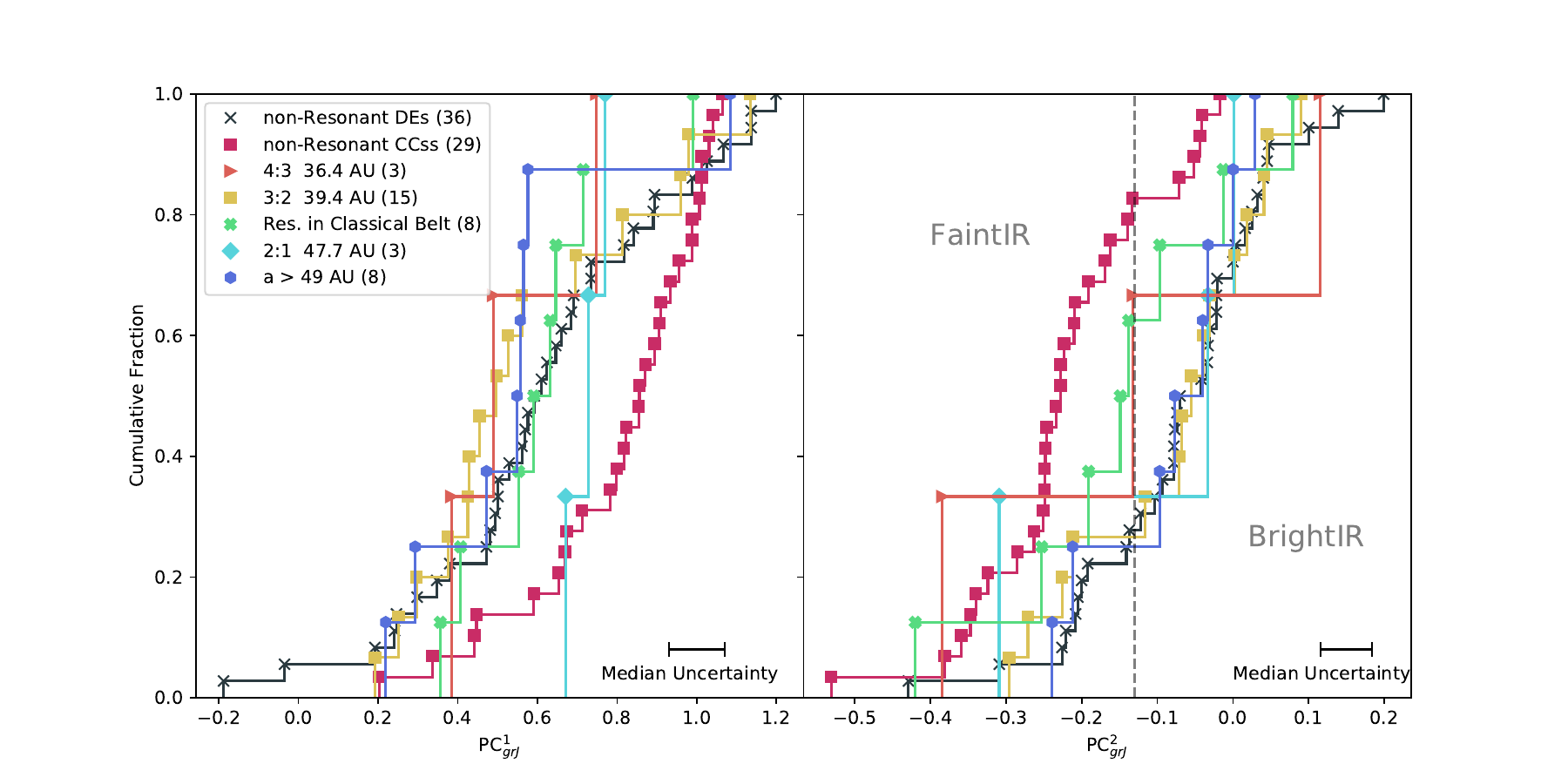}
    \caption{
    \textbf{Left:} The cumulative fraction of PC$^1_{grJ}$ values.  The non-resonant dynamically excited objects (black `x's), dynamically cold classical objects (pink squares), 4:3 resonators (orange triangles), 3:2 resonators (yellow squares), resonant objects within the classical region (green wide `x's), 2:1 resonators (cyan diamonds), and resonant objects with semi-major axes $a>49$~au (blue hexagons) are all shown separately.  
    \textbf{Right:} The cumulative fraction of Col-OSSOS targets by PC$^2_{grJ}$ values, in the same color scheme as the left plot.  The gray dashed line indicates the divide between the FaintIR (left) and BrightIR (right) surface classifications at PC$^2_{grJ}=-0.13$.  The cold classical TNOs are dominated by the FaintIR surface classification, but this surface classification is also found in the resonances, including the 4:3 resonance which is sunward of the current cold classical belt.  The non-resonant dynamically excited objects are dominated by the BrightIR surface classification, and their distribution of PC$^2_{grJ}$ is similar to the resonant $a>49$~au objects.
    \textbf{Both:}  The median uncertainty of the points is indicated in the lower right.  We resampled the data using a Gaussian distribution and the uncertainty on each point 100 times, and examined the resulting distributions.  This resulted in broadening each distribution to very close to the width of the median uncertainty.  The uncertainties on specific measurements are presented in the right panel of Figure \ref{fig:grJ}.
    } \label{fig:PCcumul}
\end{figure}

To better demonstrate the different surfaces between the different populations, Figure \ref{fig:PCcumul} shows the cumulative fraction of TNOs in each group with particular PC$^1_{grJ}$ and PC$^2_{grJ}$ values.
The cumulative PC$^2_{grJ}$ plot highlights the differences between the different populations, and also shows that small changes in the precise location of the cut between FaintIR and BrightIR (PC$^2_{grJ}=-0.13$) would change the classification of a few objects right near the division, but would not affect the overall PC$^2_{grJ}$ distribution of each dynamical classification.
The distribution of PC$^1_{grJ}$ values on the left in Figure \ref{fig:PCcumul} does show different distributions for the dynamically excited objects and cold classical objects, but is less useful for diagnostic purposes.
The value of PC$^1_{grJ}$ is related to the spectral slope an object would have if it was directly on the reddening line, so the cumulative value of PC$^1_{grJ}$ has similarities to the spectral gradient plots in previous work \citep[e.g.][]{Sheppard2012}.
The resonant object outliers in the color-color plot, particularly the 4:3 and 2:1 deep in the cold classical clump, are not at all obvious in the cumulative PC$^1_{grJ}$ distributions.

As in \citet{fraser2021}, we note that the FaintIR group (with PC$^2_{grJ}<-0.13$) contains the vast majority of the dynamically cold classical TNOs, with the exception of five to six outliers.
(One is large uncertainty and located on the division between classifications.
There is an additional outlier if the low $PC^1_{grJ}$ cold classical is also included.)
The presence of five to seven objects identified as cold classicals based on their free inclinations within the BrightIR group may simply represent the low-$i$ tail of the hotter inclination distribution of the 36 dynamically excited objects.
We tested whether this misclassification was consistent with the sample by using models of the TNO populations and the survey simulator.
For an unbiased sample with an inclination width of 14.5$^{\circ}$, which has been found to be the best fit for the hot classical region \citep{hilat}, approximately 4\% (or 1.7 objects) of the dynamically excited sample would have $i<4^{\circ}$.
We used a survey simulator \citep{lawlerFASS} to model the pointing and depth biases of our sample, and an input model distribution of cold and dynamically excited TNOs, generated by modifying the Canada-France Ecliptic Plane Survey L7 model of the hot and cold classical populations \citep{cfeps} to match the inclination widths of 14.5$^{\circ}$ \citep{hilat} for the hot population and 1.75$^{\circ}$ \citep{vanLaerhoven2019} for the cold population.
Importantly, we modeled these inclination widths as free inclination widths rather than widths relative to the invariable plane, which is a comparatively poor match to the plane of the classical Kuiper belt.
We input these modified L7 model objects until the survey simulator detected 10,000 objects, then sub-selected the sample brighter than $m_r\le$23.6, to imitate the brightness limit of Col-OSSOS, which gave a sample of $\sim$1,000 simulated detections.
When we examine the simulated observations of the intrinsically hot classical population, we find that $15\pm2$\% of the detected hot sample would have free inclinations low enough to be mis-classified as cold. 
There are 21 hot classical TNOs in our sample. 
Postulating that the five to seven BrightIR ``cold" classical TNOs are actually the low-$i$ tail of this hot population is consistent with our estimate of the mis-classification rate (5--7/28 is $\sim18-25\%$).
On the other hand, our simulated observations imply that only $1.5\pm0.5$\% of the intrinsically cold classical population have inclinations large enough to be mis-classified as hot. 
There are 22 FaintIR cold classical TNOs in our sample.
To postulate that the 5 FaintIR hot classical TNOs really belong to the intrinsically cold population would imply an inconsistently high mis-classification rate of 18\%$^{+15}_{-14}$.
Additionally, several of the FaintIR hot classicals have free inclinations far too large to be considered the high-$i$ tail of the cold population.
While these simple survey simulations based on the L7 model do not account for the full complexity of the hot and cold populations, they strongly suggest that the BrightIR interlopers in our apparently cold population are simply the low-$i$ tail of the hot population but that the FaintIR objects in the hot population are not simply the result of overlapping inclination distributions.
These FaintIR objects appear to have experienced significantly more excitation than the cold classical population.

The dynamically excited TNOs and the resonant TNOs within the FaintIR group are likely to represent objects which formed in the FaintIR region and were scattered or captured into other populations during planetary migration.
For the resonant TNOs, in the cumulative PC$^2_{grJ}$ shown on the right in Figure \ref{fig:PCcumul}, these populations show different enhancements of FaintIR surfaces relative to the dynamically excited population.
The sample of 4:3 resonators and the resonators within the classical region have a high fraction of objects with surface colors consistent with a FaintIR surface classification seen in the majority of cold classicals.
The PC$^2_{grJ}$ for two of these resonators are also among the lowest PC$^2_{grJ}$ values in the sample, reflecting the confident classification of these objects as FaintIR bodies, see Figure \ref{fig:grJ}.
The 3:2 resonance and $a>49$ resonances show a PC$^2_{grJ}$ distribution similar to the dynamically excited TNOs.
The Plutinos are the largest sub-sample, 15 objects, and are also found with the full range of surface colors.
This remains consistent with the possibility of capture from the bulk dynamically excited population, however, as that population includes $\sim20$\% FaintIR surfaces.
The inclusion of several resonant TNOs that are coincident with the bulk of the cold classical TNOs is immediately apparent, and includes a 4:3, a 2:1, and a 3:2 resonator.

The TNOs currently in Neptune's resonances include objects with a variety of origins. Some resonant objects were captured during the era of planet migration via sweeping (resonances picking up objects as the resonances move through a population already exterior to Neptune) and scattering (objects dynamically perturbed outward from the giant planet region that end up in Neptune's resonances); there are also objects from the current scattering population that temporarily stick in the resonances. Depending on the specifics of planetary migration, different resonances may have preferentially captured objects from different primordial planetesimal populations (we expand upon this in Section~\ref{sec:Discussion}).
It is generally thought that the dynamically excited populations originate from a portion of the planetesimal disk closer to the giant planets than the cold classical TNOs thought to have formed in situ (see review by \citealt{Gladman2021}), which is motivated by, and consistent with, the idea of a color transition in the planetesimal disk, as discussed above.
In this scenario, we expect the resonances that are dominated by capture of scattering objects (either in the early solar system or from the current scattering population) to have surface properties similar to the dynamically excited population; while resonances dominated by sweeping capture of cold disk objects should share surface properties with the cold classical population.

We used two different methodologies for comparing the different sub-populations: (1) an examination of the apparent distribution, in Section \ref{sec:apparent} and (2) a survey simulation on a model to determine the intrinsic ratios, in Section \ref{sec:intrinsic}.
For (1), we look at the apparent distribution of the objects in terms of their $PC^2_{grJ}$ values.
The apparent surface distribution is a good proxy for the intrinsic distribution \textit{if} the absolute magnitude $H$-distribution of the FaintIR and BrightIR objects is similar.
Recent work by \citet{petit2023} and \citet{kavelaars2021} find that an exponential taper shape provides a good fit for the size distribution for the hot and cold main belts respectively.
In the range 5.8$<H<$8.3, \citet{petit2023} find that the same exponential taper size distribution shape can be used to model both the hot and cold populations.
This is different result than previous works which used multi-component power laws, and found different parameters for the hot and cold populations \citep[e.g.][]{cfeps,fraser2014}.
The 5.8$<H<$8.3 absolute magnitude range includes the overwhelming majority of the Col-OSSOS sample excluding the Centaurs, so if the similarly shaped size distribution of \citet{petit2023} $H$-distribution is accurate then the apparent distribution of surfaces in our sample can be interpreted as representative of the intrinsic distribution.
The comparison of apparent surface property distribution has the advantage of not being dependent on orbit distribution models of different sub-populations, however, if the $H$-distributions are dependent on surface-type, it may not be robust for comparing sub-populations with very different discovery biases.
For (2), we use a model of the TNOs and a survey simulator to replicate the survey biases and determine what intrinsic surface type fractions are consistent with the observations.
This methodology has the advantage that different $H$-distributions can be used for different sub-populations, however, the results rely on having a good intrinsic model of the orbital distribution of the TNOs, and we treat the classification of surface type as binary instead of a distribution of $PC^2_{grJ}$ values.
We discuss the orbital model and survey simulations used in Section \ref{sec:intrinsic}.
The combination of these two approaches provides useful insight into the distribution of surfaces in the outer solar system.

\subsection{Apparent Surface Classification Distribution}
\label{sec:apparent}

In order to determine whether the surface classifications of resonant TNOs are consistent with capture from the current dynamically excited population, we used the Anderson-Darling \citep[AD,][]{AndersonDarling} statistical test, which is similar in concept to the often used Kolmogorov–Smirnov (KS) test. The AD test statistic is a measure of the difference between the cumulative distributions of a parameter from two populations and can be used to assess the probability that the two populations are sub-samples drawn from the same parent population.
The significance of the AD statistic is computed with a bootstrapping method, which re-samples the assumed parent population distribution to compare against itself to compute the expected distribution of the AD statistic for the case where we know the two samples are related.
For the resonant objects, we consider the possibility that the parent population is either: (1) the dynamically excited TNOs, (2) the cold classical TNOs, or (3) a mixture of the two.

\begin{figure}[h!]
    \includegraphics[width=.9\textwidth]{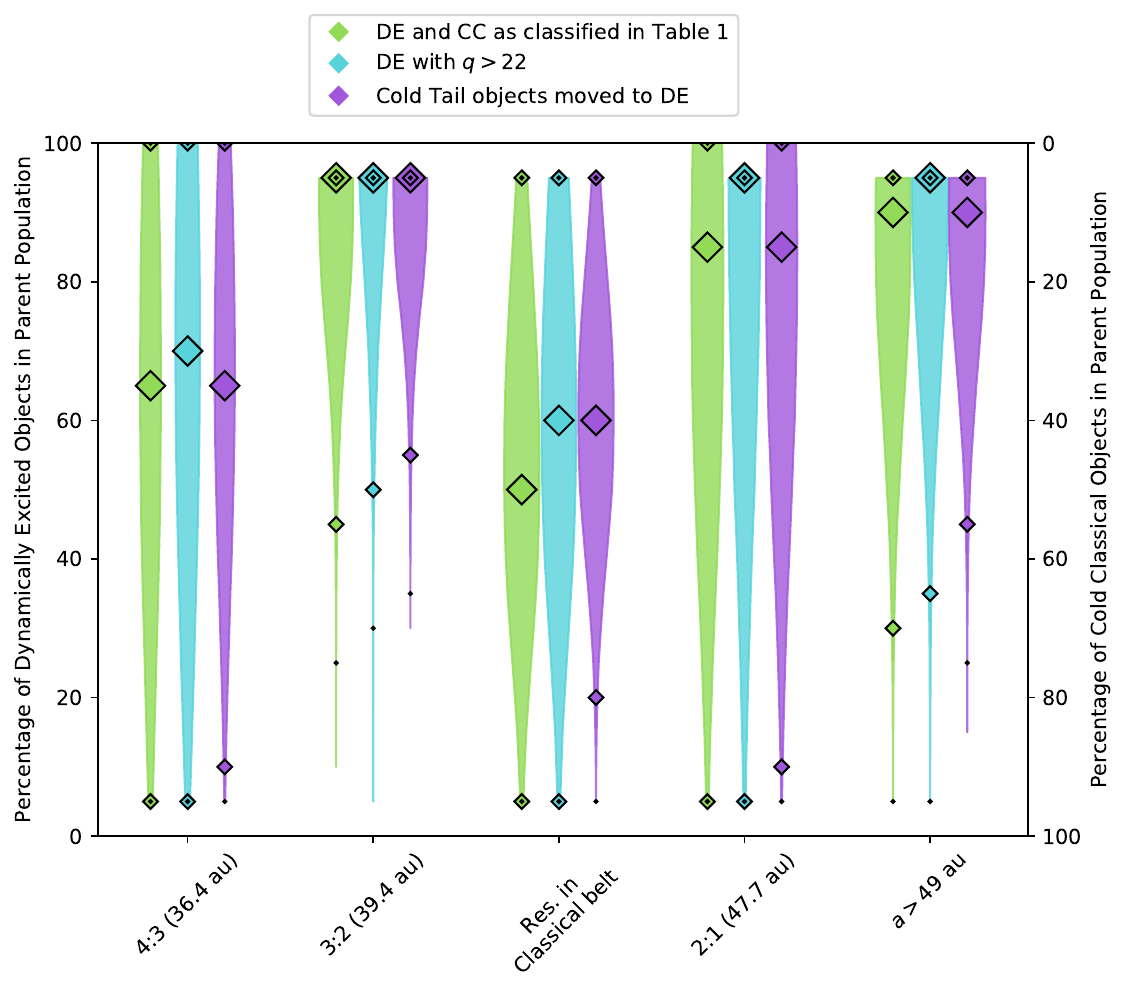}
    \caption{The results are presented for the apparent parent populations for the different resonant populations.  The different resonances are indicated along the x-axis.  The parent populations are a combination of the dynamically excited TNOs (DE) and cold classical TNOs (CC), with different percentages of each of the two groups, from 0--100\% calculated in steps of 5\%.  The distance between the distribution of resonant PC$_{grJ}^2$ values and parent population PC$_{grJ}^2$ values is computed with the AD test, then the significance of that result is obtained by bootstrapping using the input parent population.  The large diamonds indicate the least rejectable D-statistic parent population fraction, the small diamonds indicate the 2$\sigma$ limits, and the smallest (closed) diamonds indicate the 3$\sigma$ limits based on the D-statistic.  The width of each bar is proportional to the D-statistic value at each 5\% step.  The three different bars for each resonance show the results for slightly different assumptions in the parent populations.  The green (left) bar uses the DE and CC populations exactly as defined in Table \ref{tab:targets}.  The teal (middle) bar uses the CC objects as defined in Table \ref{tab:targets}, but for the DE objects uses only the DE objects with pericenter $q>22$.  The purple (right) bar uses the CC and DE objects as classified in Table \ref{tab:targets} except that the 5 cold classicals with $PC^2_{grJ}$>-0.13 BrightIR surfaces are moved from the CC group to the DE group, as 5 low-$i$ objects is consistent with being the low-$i$ tail of the DE inclination distribution, see Section \ref{sec:apparent} for details.  Note that the Parent Population used here is to reproduce the \textit{apparent} surface distribution and not the \textit{intrinsic} population ratios of the dynamically excited and cold classical objects- due to detection biases and albedo differences, see main text for details.
    } \label{fig:ADstats}
\end{figure}

To determine what source population is consistent with the resonant TNO PC$_{grJ}^2$ values, we compared each of the apparent resonant PC$_{grJ}^2$ distributions with: the apparent PC$_{grJ}^2$ distribution of the dynamically excited TNOs; the apparent PC$_{grJ}^2$ distribution of the cold classical TNOs, and source populations which are a mixture of the apparent distribution of cold classical and dynamically excited TNOs.
We created possible parent populations by combing the two possible parent populations (dynamically excited and cold classical TNOs) with different relative contributions of the two components, ranging from 0-100\% in steps of 5\%.
We cloned the full list of PC$_{grJ}^2$ values for each of components until they reached the desired total ratio.
Possible parent populations for the resonant TNOs include the 100\% dynamically excited TNOs, 95\% dynamically excited TNOs with 5\% cold classical TNOs, etc.
In Figure \ref{fig:ADstats}, the results of the comparisons between the different resonant populations and possible source populations are given.
The values within the 2$\sigma$ limits are parent population models which are least rejectable, indicated by small diamonds.
We find, unsurprisingly given the known optical color distribution, that all of the resonances reject at $>3\sigma$ a scenario where their source population is made up entirely of the cold classical population.
However, a wide range of mixed contributions from the two populations were non-rejectable.

We also investigated two additional variations on the parent populations: excluding the Centaurs and moving the five dynamically cold classical objects with high $PC^2_{grJ}$ values (which we previously argued are consistent with the low-$i$ tail of the dynamically excited population) into the dynamically excited population.
There are several Centaurs in the Col-OSSOS sample, which are classified as dynamically excited TNOs.
Because the surface properties of Centaurs with small perihelion distances can potentially have been altered since formation by thermal modification in the Centaur region \citep[e.g.][]{Jewitt2009}, we also tested the sub-population of dynamically excited TNOs with pericenter $q>22$; the exclusion of these objects did not significantly affect the results.
We also repeated the AD tests with the five $PC^2_{grJ}>$-0.13 BrightIR objects with cold classical orbits moved to the dynamically excited objects lists.
All three scenarios, (1) the objects exactly as classified in Figure \ref{fig:aei} and Table \ref{tab:targets}; (2) dynamically excited objects limited to $q>22$; and (3) low-$i$ tail of dynamically excited objects moved to dynamically excited sample, are included in Figure \ref{fig:ADstats} for comparison.

The lower-$a$ resonances with the exception of the 3:2 resonators do show some trends toward larger contributions from the cold classical population.
The `least rejectable' parent populations for the 4:3 (interior to the classical belt at 36.4\,au), 2:1 (beyond the classical belt at 47.7\,au), and resonances within the classical belt include a significant contribution from the cold classical population (30-50\%).
Additionally, the PC$_{grJ}^2$ values of the objects in the 4:3 resonance and the resonant within the classical belt population reject the hypothesis that they are sourced exclusively from the dynamically excited population.

\subsection{Intrinsic Surface Classification Distribution}
\label{sec:intrinsic}

As the Col-OSSOS sample is essentially complete, we are able to robustly test models of the intrinsic distribution of objects using a model of the TNO orbit and size distribution as well as a survey simulator.
We give a brief description of the survey simulator here; see \citet{lawlerFASS} for a more detailed explanation on its use.
We utilize the OSSOS survey simulator \citep{lawlerFASS}, which takes as input the pointings and characterization of the survey pointings as well as a model of the TNO distribution, and returns a list of synthetic detections which have been biased in the same way as the real detections.
The real detections can then be compared to the synthetic detections to determine whether the intrinsic model is consistent with the real detections.

We modified the survey simulator input characterization to reflect the Col-OSSOS sample.
This included restricting the pointings to OSSOS E, H, L, O, S, and T blocks and rejecting any $a<30$ objects `detected' in O block, as the unmeasured Centaur is in O block.
We also modified the survey simulator to only `count' synthetic detections with $m_r<23.6$, as the Col-OSSOS sub-sample of OSSOS is objects composed of objects brighter than this limit.

As input for the survey simulator, we used a modified version of an OSSOS model of the TNO orbital and size distribution, which is based on the TNOs in \citet{bannister2018}.
This model of the TNOs includes all of the sub-populations present in the OSSOS++ surveys \citep{cfeps,hilat,alexandersen2016,bannister2018}, and uses orbital distribution and $H$-distribution models of the sub-populations derived from publications related to those surveys.
There are some minor tweaks to the parametric distributions, within the error bounds of the previously published models, to improve the match between the biased model detections and the OSSOS++ detections.
The model includes objects $H_r<12.0$ for all sub-populations.
The model cold classical TNOs use the debiased orbital distribution of the OSSOS++ cold classicals, and is extended to smaller sizes using the CFEPS L7 model size distribution \citep{cfeps}.
The model hot classical TNOs use the debiased orbital distribution from OSSOS++, and the extended classical belt (sometimes referred to as detached objects) uses the orbital distribution from the CFEPS L7 model.
The model hot classicals, extended classical belt, scattering objects, and resonant populations all use the power-law based size distribution from \citet{greenstreet2019}.
The model scattering objects use the $a-e-i$ distribution of scattering objects from \citet{kaib2011_ICARUS,kaib2011_ApJ}.
The orbital distributions of the model resonant TNOs were taken from several publications for all resonances with two or more detections.
The 3:2 distribution is from \citet{volk2016}, the 2:1 is from \citet{chen2019}, and the other resonant population orbital distributions are from \citep{crompvoets2022}.
We separated the OSSOS TNO model into sub-populations corresponding to our real object sub-populations: cold classical, dynamically excited, 4:3, 3:2, resonant within the classical belt, and resonant a>49~au.
With this methodology, the assumption is that the FaintIR objects have the same size distribution as the cold classical objects, and the BrightIR objects have the size distribution of the dynamically excited objects.
We produced two models of each sub-population, one using the size distribution from the cold classicals and one using the size distribution from the dynamically excited objects from the OSSOS TNO model.
This resulted in 14 sub-population models to test with the survey simulator.

\begin{figure}[h!]
    \includegraphics[width=.9\textwidth]{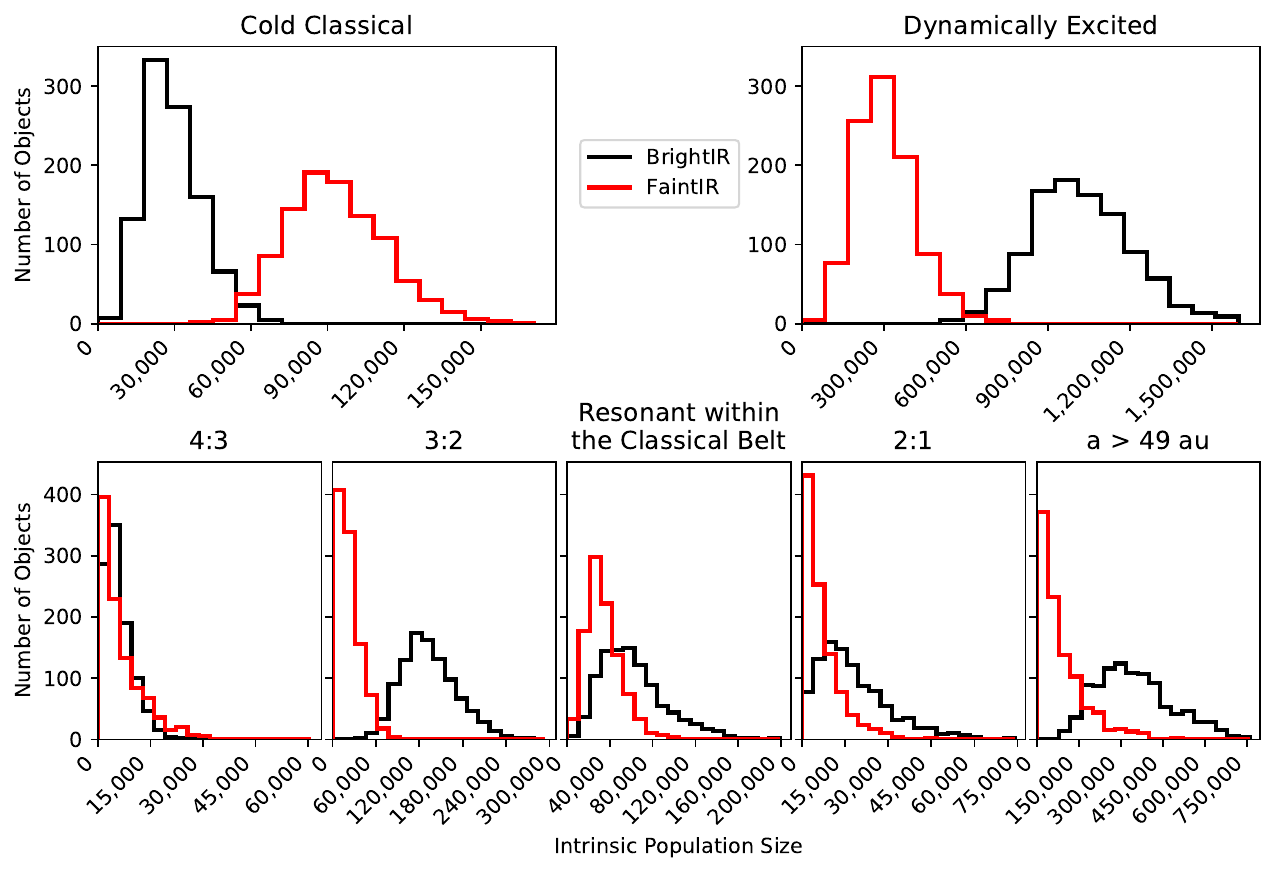}
    \caption{The population estimates of the intrinsic population required to reproduce the number of BrightIR and FaintIR detections found by Col-OSSOS in each sub-population.  The input model is for $H_r<12$ and is based on the OSSOS TNO model, see main text for details.  The relative intrinsic population sizes required to produce the number of real Col-OSSOS detections vary significantly for different sub-populations.  The 3:2 distribution and the resonant $a>49$~au distribution are quite similar to the dynamically excited distribution, while the other three resonant sub-populations are in between the cold classical and dynamically excited distribution, suggesting significant contributions to those resonances from both possible parent populations.
    } \label{fig:popest}
\end{figure}

The survey simulator can be used for a few different kinds of analysis, and we utilized it to determine an intrinsic population estimate.
The resulting constraints on the number of FaintIR and BrightIR objects in each sub-population are listed in Table \ref{tab:popEst}.
For the FaintIR and BrightIR size distributions with each sub-population, we ran the survey simulator to see how many intrinsic objects were necessary to produce the number of Col-OSSOS objects with FaintIR and BrightIR surfaces respectively.
To do this, we used the two models of each sub-population, which differed only in their $H$-distributions: one using a dynamically excited $H$-distribution and the other using a cold classical $H$-distribution. 
We ran the survey simulator until the expected number of FaintIR or BrightIR objects was detected, depending on the input model.
(For example, for the 3:2, we created two input 3:2 models: the 3:2 orbital distribution with a cold classical $H$-distribution and the 3:2 orbital distribution with the dynamically excited $H$-distribution.
We provided both of those input models to the survey simulator separately, and calculated how many intrinsic objects were required to produce 2 detections for the cold $H$-distribution and 13 detections for the excited $H$-distribution.)
This was repeated with a different random seed 1000 times for each input model to determine the intrinsic population estimate for objects in each sub-population with $H_r<12$.
The results are reported in Figure \ref{fig:popest}, which shows the distribution of intrinsic population sizes which produce the appropriate number of simulated detections.
Table \ref{tab:popEst} also reports the median and two sigma values for the population estimates, as `Intrinsic FIR' and `Intrinsic BIR'.
The ratio of the median intrinsic FIR/BIR populations is also reported in this table; refer to Figure \ref{fig:popest} and the population estimate 2$\sigma$ uncertainties for an estimate of the uncertainty on this number.
The apparent ratio of FIR/BIR is also provided in Table \ref{tab:popEst} for comparison.

\startlongtable
\begin{deluxetable}{l l c c c c c c c c c c c}
\tabletypesize{\scriptsize}
\tablecaption{\label{tab:popEst}
Sub-population FaintIR (FIR) and BrightIR (BIR) Surface Types: Apparent and Intrinsic Ratios.  The number of each surface type in the Col-OSSOS sample and the basic ratio of FIR/BIR this represents is given.  The intrinsic FIR and BIR numbers are the median and 2$\sigma$ population estimate values calculated using an input model and the survey simulator, see Section \ref{sec:intrinsic} for details.  The ratio of intrinsic FIR/BIR is calculated directly from the median population estimates.
}
\tablehead{ 
\colhead{Population}  & \colhead{Number FIR} & \colhead{Number BIR} & \colhead{Apparent FIR/BIR Ratio} & \colhead{Intrinsic FIR}& \colhead{Intrinsic BIR} & \colhead{Intrinsic FIR/BIR Ratio}  
}
\decimals
\startdata
4:3 & 1 & 2 & 0.5 & 4,400$^{+16,000}_{-4,000}$ & 5,100$^{+8,800}_{-4,200}$ & 0.85 \\
3:2 & 2 & 13 & 0.15 & 19,000$^{+36,000}_{-15,000}$ & 130,000$^{+73,000}_{-51,000}$ & 0.15 \\
Res. In Classical Belt & 4 & 4 & 1 & 31,000$^{+35,000}_{-19,000}$ & 56,000$^{+68,000}_{-34,000}$ & 0.55 \\
2:1 & 1 & 2 & 0.5 & 4,700$^{+16,000}_{-4,300}$ & 15,000$^{+29,000}_{-12,000}$ & 0.31 \\
a$>$49 au & 1 & 7 & 0.14 & 60,000$^{+207,000}_{-55,000}$ & 330,000$^{+280,000}_{-170,400}$ & 0.18 \\
Dynamically Excited & 8 & 28 & 0.29 & 290,000$^{+210,000}_{-140,000}$ & 1,000,000$^{+340,000}_{-270,000}$ & 0.29 \\
Cold Classical & 22 & 7 & 3.14 & 92,000$^{+36,000}_{-28,000}$ & 28,000$^{+22,000}_{-14,000}$ & 3.28 \\
\enddata
\end{deluxetable}

The intrinsic FIR/BIR ratio determined through survey simulations is model dependent, however, we have selected a model which is consistent with the 800+ discoveries from which the Col-OSSOS sample was sub-selected.
Similar but slightly different intrinsic ratios would result from other models consistent with the OSSOS++ detections, and the results are most sensitive to changes in the size distributions.
In Table \ref{tab:popEst}, it is clear that the intrinsic population ratios of FIR/BIR objects are sometimes quite similar to the apparent ratios (for example, for the dynamically excited objects) and sometimes noticeably different (for example, for the resonant objects within the classical belt).
Overall, differences between the apparent and intrinsic ratios do not significantly affect our interpretation of the results.
The 3:2 and $a>49$~au population estimates in Figure \ref{fig:popest} very closely resemble the dynamically excited population estimate distribution and the other resonances (4:3, 2:1, and resonant within the classical belt) all have population estimate distributions which appear to be a blend of the cold classical and dynamically excited populations.

\section{Discussion}
\label{sec:Discussion}

The resonant populations preserve a record of the capture events that occurred during planetary migration.
As Neptune migrated, its resonances trapped outwardly scattered objects \citep[e.g.][]{gomes2003} as well as dynamically colder objects that the resonances swept \citep[e.g.][]{malhotra1995} during migration.
Present-day scattering objects can also become transiently stuck to resonances as their orbits evolve \citep{lykawkamukai07}.
These different capture mechanisms have different source regions in the original planetesimal disk.
The scattering capture which occurs during planetary migration would likely have caught objects from the primordial disk which were scattered outward from the region currently occupied by the giant planets \citep[e.g.][]{Gomes2005,levison2008}; see also review by \citealt{Nesvorny2018}. 
This differs from modern transient resonance sticking from the scattering population in that the still on-going migration allowed these objects to evolve into more stable regions of the resonances.
The precise efficiency of this capture depends on the initial conditions of the disk and the specifics of the planetary migration scenario \citep[e.g.][]{murray-clay2006,nesvorny2015b,volkmalhotra2019}.
These objects scattered and captured into resonance during planetary migration are likely to have similar origins to the bulk dynamically excited Kuiper belt region \citep[e.g.][]{levison2008}, which include TNOs currently in the hot classical, detached, and scattering populations.
The current, transiently sticking resonant objects represent captures from the reservoir of currently scattering TNOs; while not all transient resonant captures are particularly recent (residency timescales are roughly evenly distributed in log time; \citealt{Yu2018}), they are sourced from the post-migration scattering population and thus share an origin with the dynamically excited population.
Thus resonant objects should have the same surface distribution as the dynamically excited TNOs, i.e. dominated by BrightIR surface classifications, if all of the resonant objects are captured through scattering capture during migration and present-day transient resonance sticking.
However, if Neptune's migration included a phase of smooth migration, it may have swept TNOs from the current cold classical region, thought to be a preserved remnant of the in situ primordial disk, into some of the resonances.
Sweeping capture into the resonances during migration is orders of magnitude more efficient than scattering implantation \citep[e.g.][]{volkmalhotra2019}, so if sweeping capture through the primordial cold classical belt occurred for a resonance, a large population of FaintIR surfaces is expected.
While smooth migration alone does not adequately explain all of the observed TNO populations, Neptune might have smoothly migrated outward at the end of its migration \citep[see, e.g.][]{nesvorny2015b,nesvorny2015a}, allowing resonances in the current classical belt region to preferentially capture objects from the primordial cold classical population, dominated by objects with the FaintIR surface classification.

\begin{figure}[h!]
    \includegraphics[width=1.\textwidth]{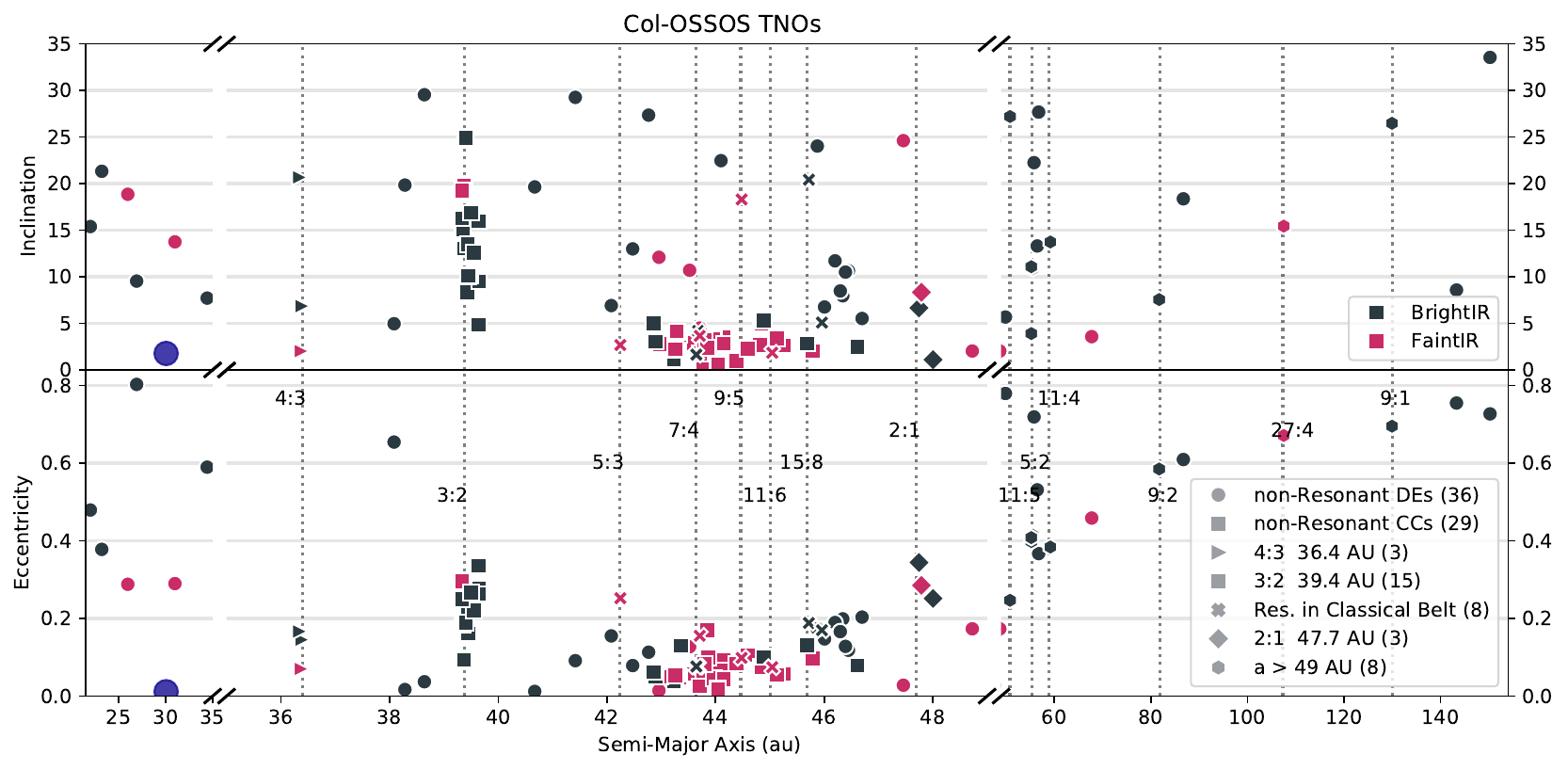}
    \caption{The orbital distribution of Col-OSSOS TNOs, with surface classification indicated.  As in Figure \ref{fig:aei}, the semi-major axis is scaled by three different amounts, as indicated in the x-axis and separated by gaps, and the symbol shapes are preserved, but the colors of the points indicate BrightIR classification (black) and FaintIR classification (red).  The low-inclination and low-eccentricity 4:3 resonator at 36.4~au is particularly evident, as is the classification-inclination dependence, particularly at large-$a$, noted by \citet{marsset2019, marsset2022}.  Neptune is indicated by the large blue circle.
    } \label{fig:aei_surfaceClass}
\end{figure}

In this work, we find some FaintIR surface classifications in all dynamically excited orbital classifications, at varying degrees of representation.
The distribution of FaintIR and BrightIR surfaces based on their orbital parameters is shown in Figure \ref{fig:aei_surfaceClass}.
Arguably the most dynamically interesting FaintIR surface is found in the 4:3 resonance.
This object, 2013 US$_{15}$ is optically very red and falls conclusively within the FaintIR surface classification group.
2013 US$_{15}$ has a semi-major axis of 36.38~au, sunward of the current cold classical region.
2013 US$_{15}$ is also the least dynamically excited 4:3 resonator in the sample, with $e=0.07$ and $i=2.02^{\circ}$, similar to the cold classical orbits, but at smaller semi-major axis.
The inclusion of this object into the 4:3 resonance requires further investigation.
A straightforward explanation for the presence of a FaintIR 4:3 resonator is that the primordial FaintIR surface type extended sunward of the current cold classical belt to at least 36~au, and some resonance sweeping occurred to trap this object into the 4:3 resonance.
The low excitation of its current orbit (low $e$ and low $i$) implies it could not have been carried very far from its original location by the resonance.
To test the stability of this object, we integrated 250 clones sampled from the orbital uncertainties for $>$3~Gyr.
The orbits of the clones were all stable for the duration of the integration, and the libration amplitude of the resonator was also stable at 94$^{\circ}\pm$3$^{\circ}$ (comparable to its current libration amplitude of 95$^{\circ}\pm$1$^{\circ}$ determined from a 10 Myr integration).
Resonance sticking with similar libration amplitudes is possible, but not common \citep{Yu2018}, and the long-term stability of the entire orbital cloud supports a formation in-situ and a deep, primordial capture into the 4:3 resonance.
Currently, the FaintIR surface of 2013 US$_{15}$ accounts for one third of the sample of 4:3 resonators, so the fraction of the 4:3 resonators with FaintIR surfaces may be quite significant.
While the intrinsic FaintIR/BrightIR surface ratio is difficult to constrain with this small sample, we emphasize that the presence of a single object which is inconsistent with scattering capture is strong evidence that the FaintIR surface type formed significantly inward of the current cold classical belt whether or not the high fraction of FaintIR surfaces holds with a larger sample.
However, additional optical and near-infrared observations of 4:3 resonators would be extremely helpful for understanding the intrinsic frequency of FaintIR surfaces in this resonance.

Just sunward of the cold classical belt, the 3:2 resonance includes some FaintIR objects.
However, whether one considers the apparent or intrinsic surface classification fractions, the fraction of FaintIR surfaces within the 3:2 resonance is more similar to the non-resonant dynamically excited TNOs than other resonances near and inside the cold classical belt.
This was also noted for the apparent optical color distribution of 3:2 resonators in the literature examined by \citet{jewitt2018}.
In Figure \ref{fig:ADstats}, it is clear that the 3:2 resonance rejects a significant contribution from the cold classical population, and favors a higher contribution from the Dynamically Excited population, $90^{+5}_{-40}$\%.
In the intrinsic FaintIR/BrightIR ratios in Table \ref{tab:popEst}, the 3:2 may even have a smaller fraction of FaintIR surfaces than the bulk dynamically excited population.
While this reduced fraction of FaintIR surfaces in the 3:2 \textit{could} indicate a lack of resonance sweeping and/or that the primordial FaintIR objects did not form in the region of the current 3:2, the secular resonances near the 3:2 and the orbital dynamics of the resonators complicate the interpretation of the small number of FaintIR surfaces in the 3:2 resonance.
In other resonances, the FaintIR objects were predominantly those with lower-$i$ orbits, whereas the OSSOS sample of 3:2 objects examined here have moderate- to large-$i$, $i\gtrsim$5$^{\circ}$.
Low-inclination 3:2s are not common, but have been found in other surveys \citep[e.g.][]{alexandersen2016}, so these objects warrant additional observations to determine whether their surfaces fall preferentially into the FaintIR category.
We also note that resonant sweeping (if it occurred) might have been less effective near the 3:2 resonance if the intrinsically cold population sunward of the 3:2 was cleared by secular resonances during planetary migration before the 3:2 could sweep it. 
The inner edge of the modern-day cold classical population is sculpted by the $\nu_8$ eccentricity secular resonance which destabilizes low-inclination orbits in the 40-42~au region adjacent to the 3:2 resonance (see, e.g., discussion in \citealt{Gladman2021}), which might have reduced the capture efficiency of the 3:2 during sweeping; there is also an inclination secular resonance in the same region that could excite the inclinations of initially cold objects \citep[e.g.][]{Chiang2008}. 
The locations and strengths of these secular resonances during planet migration could have significantly affected capture probabilities for cold objects into the 3:2 and their final inclinations if captured (see discussion in \citealt{volkmalhotra2019}).
Based on the 3:2 sample here, we do not find evidence of resonant sweeping capture into the 3:2, but the complex dynamics near this resonance and the mid- to high-inclinations of the observed sample do not provide conclusive evidence against sweeping having occurred.

The Col-OSSOS sample has some additional FaintIR objects of note, including objects which are resonant within the cold classical belt (in the 5:3, 7:4, 9:5, 11:6, and 15:8 resonances).
These resonances, which overlap in semi-major axis with the current cold classical region, have a significantly enhanced fraction of FaintIR surfaces seen in both the apparent and intrinsic surface classification fractions.
The 7:4 and 11:6 resonators are both low-$e$ and low-$i$ $(<$4$^{\circ}$), the 5:3 object is low-$i$ but large-$e$, while the 9:5 resonator has an excited orbit with $i=$18.3$^{\circ}$.
The higher-$i$ orbit of the FaintIR 9:5 may be the result of diffusion within the resonance, as this object is in a mixed $e$-$i$ resonance.
The enhancement in FaintIR surfaces for the resonances within the classical region is seen in Figure \ref{fig:PCcumul} and Figure \ref{fig:ADstats}, and the preferred source population is 60\% dynamically excited TNOs and 40\% cold classicals (see Figure \ref{fig:ADstats}).
In the intrinsic population analysis for the resonances within the classical belt, a large component of FaintIR surfaces is also required in order to produce the observed surface distribution.
The enhancement of the FaintIR surface class in the resonances within the cold classical region also supports the conclusion that there was some amount of resonant sweeping which captured FaintIR surfaces into these resonances.

Just beyond the cold classical belt, the 2:1 resonance includes a FaintIR object.
This object, 2013 JE$_{64}$, is firmly within the FaintIR surface classification, and has an excited orbit ($e=0.29$ and $i=$8.34$^{\circ}$).
The larger $e$ and $i$ may suggest a larger distance of resonance sweeping before capture, or that this particular object was captured through scattering/resonance sticking.
As with the 4:3 resonance, a larger sample of optical and near-infrared surface colors of 2:1 resonators is needed to better constrain the intrinsic fraction of FaintIR and BrightIR surface classifications.

The distant $a>49$ resonances include the smallest apparent fraction of FaintIR surfaces of the resonances and the 3:2 and $a>49$ resonances have the lowest intrinsic fraction estimates.
The surface class distribution is very similar to the non-resonant dynamically excited surfaces shown in Figure \ref{fig:PCcumul} and the intrinsic population estimate distribution is similar to the dynamically excited population estimate distribution in Figure \ref{fig:popest}.
This is consistent with the FaintIR objects in the outer resonances being captured via scattering during planetary migration and/or current scattering capture.
We note that, similar to the 3:2, the sample does not include $i<5^{\circ}$ distant resonators.
Significant numbers of low-$i$ resonators are not expected in these distant resonances, which have intrinsically hot inclination distributions \citep[e.g.][]{volk2016, pike2017}.
We also do not necessarily expect these more distant resonances to have had a cold, primordial FaintIR disk population to have swept during any period of smooth migration.
The currently known cold population does not extend past the 2:1, so Neptune's end-stage smooth migration would need to have extended for at least 4 au for the next major resonance, the 5:2, to have had any chance of sweeping up objects from the known cold population.
If low-$i$ resonators are found in these distant resonances, particularly the 5:2, measuring their surface properties in the optical and near-infrared would be helpful in determining the primordial outer limit of the disk of FaintIR objects.

The objects in this sample come from a large program which attempted to acquire a brightness complete sub-sample of several blocks of the OSSOS discoveries.
The overwhelming majority of the targets which met this criteria, 102 of 103, have $g-r$ and $r-J$ surface colors acquired as part of the Col-OSSOS project.
The un-observed target could not be measured because it had moved into the galactic plane since discovery, and was not omitted or unmeasured due to brightness or data quality issues.
We do not expect this missing object to affect our conclusions in any way, and have omitted any synthetic detections of O-block Centaurs (like this object) from our survey simulations to correct for this unmeasured object.

\begin{figure}[h!]
    \includegraphics[width=.9\textwidth]{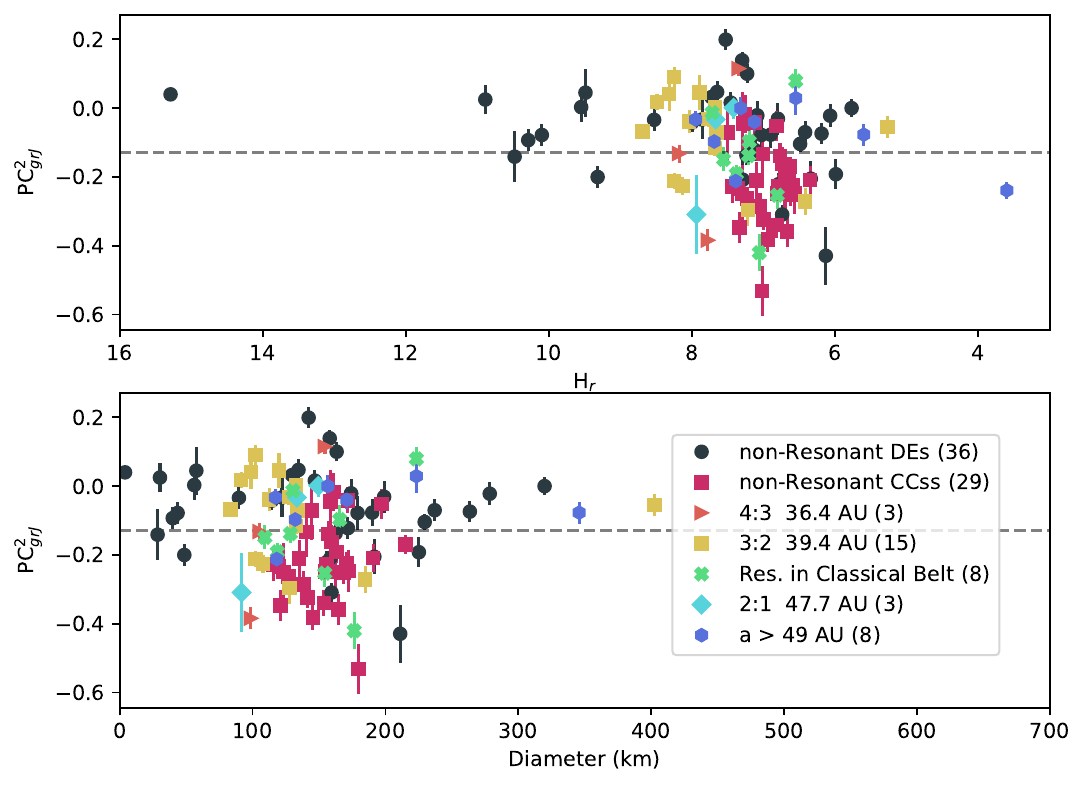}
    \caption{\textbf{Upper:} The PC$^2_{grJ}$ values and absolute magnitude in $r$-band $H_r$ at discovery.  The dynamically excited objects span a large range of values, and the fraction above and below the PC$^2_{grJ}$ division does not obviously depend on $H_r$.  The different dynamical classifications are consistent with previous figures: non-resonant Dynamically excited objects (black circles), dynamically cold classical objects (pink squares), 4:3 resonators (orange triangles), 3:2 resonators (yellow squares), resonant objects within the classical region (green wide `x's), 2:1 resonators (cyan diamonds), and resonant objects with semi-major axes $a>49$~au (blue hexagons). \textbf{Lower:} The distribution  PC$^2_{grJ}$ values compared to object size, assuming albedos of 0.14 for PC$^2_{grJ}<-0.13$ and 0.085 for PC$^2_{grJ}>-0.13$.  The distribution is shifted slightly compared to the $H_r$ panel, and the result is that the Plutinos and cold classicals appear to be more comparable in terms of intrinsic size.
    } \label{fig:size}
\end{figure}

The Col-OSSOS sample was selected to be brightness-complete based on the $r$-band discovery magnitude, $m_r\le23.6$, which imposes a specific bias on the sample.
Because the cut is made in $r$-band, the size limit for red and very red objects is different; the sample would include smaller very red objects, as a similar sized red object would have an $r$-band magnitude fainter than the cut.
The solar system absolute magnitude at discovery is shown in Figure \ref{fig:size} compared to the PC$^2_{grJ}$ values.
Six of the seven smallest objects have the BrightIR surface classification.
It is possible that this difference in surface classification is based on size, however, the majority of non-resonant faint dynamically excited objects are also BrightIR surfaces.
As we noted in Section \ref{sec:surfaces}, excluding the objects with pericenters $q<22$~au (whose surfaces may have evolved due to cometary-like activity) did not affect the results of our analysis.
We note that the non-resonant dynamically excited TNOs span the range of $H_r$ magnitudes seen by the majority of the resonant objects and the cold classicals.

A main prediction of \citet{fraser2021} was that the FaintIR group would share similar properties, such as higher albedo than the BrightIR group.
To understand how this would affect sample characteristics, we calculated the diameter of the objects assuming albedos of 0.14 for BrightIR and 0.085 for FaintIR surfaces (PC$^1_{grJ}>0.4$ and PC$^2_{grJ}>-0.13$) based on typical values for hot classical and cold classical TNOs from \citet{vilenius2018}.
Using these albedo assumptions we find that the diameter distribution of non-resonant dynamically excited TNOs spans a large range of diameters, $D$, but there is no obvious dependence between surface classification and size in the size range of the Col-OSSOS sample, see Figure \ref{fig:size}.
The size range of the Plutinos and cold classical TNOs is more similar than their $H_r$-magnitude range.
Based on the lack of obvious correlations in the distributions of $H_r$ and $D$, we conclude that it is appropriate to directly compare the distribution of surfaces within the different dynamical classifications.

The distribution of FaintIR and BrightIR surfaces within Neptune's resonances appears dependent on semi-major axis.
The FaintIR and BrightIR classification system is more robust than using purely optical surface colors for classification.
The color fractions of the Neptune resonances provide a useful criteria for assessing different models of planetary migration and constraining the initial distribution of surface classifications in the primordial disk and motivates work similar to  \citet{nesvorny2020} and \citet{buchanan2022}, but using the new surface classifications.
Carefully selected targets representative of the Col-OSSOS sample would make excellent targets for spectroscopy using the James Webb Space Telescope, and we expect future spectra of TNOs to identify the link between surface composition, surface classification, and formation location.

Constraining the characteristics of the initial proto-planetesimal disk is significant motivator for Col-OSSOS and other TNO surface studies.
Our preferred explanation for the inclusion of a FaintIR surface, also seen for the bulk of the cold classicals, in the 4:3 resonance is that the original extent of the FaintIR objects began sunward of the current 4:3 resonance (36.4~au).
The higher fraction of FaintIR surfaces in the resonances within the classical belt and the inclusion of a FaintIR object in the 2:1 resonance implies that this primordial FaintIR surface extended through the current cold classical region and at least close to the 2:1 resonance.
The high fraction of FaintIR surfaces in these resonances means that at least some small amount of sweeping migration was likely, however, a larger sample of 4:3 and 2:1 surface colors in optical and near-infrared would provide a more robust measurement of the intrinsic color fraction in these resonances.
The reduced fraction of FaintIR surfaces in the 3:2 is less clear to interpret due to both the potential influence of secular resonances on capture probabilities during migration (smooth or otherwise) and the lack of low-$i$ 3:2s in the Col-OSSOS sample.
The distant resonances ($a>49$~au) do not show evidence of an enhancement of FaintIR surfaces compared to the wider dynamically excited population, but a larger sample of surface properties for objects in the 5:2 resonance would provide a useful constraint on the limit of the outer extent of the primordial FaintIR surface objects.
New planetary migration simulation work is needed to determine the range of planetary migration scenarios and initial disk conditions (particularly the original extent of the cold classical disk and the effect of secular resonances on this cold disk) which would reproduce the surface classification distribution measured in Col-OSSOS.

Expanding the sample of objects in the 4:3, 2:1, and 5:2 resonances with high signal to noise near-simultaneous optical and near-infrared photometry is critical to provide more strict constraints on the intrinsic population fractions of the resonances.
Some additional targets from current/past large surveys could be used to expand this sample, however, future TNO discoveries by the Vera Rubin Legacy Survey of Space and Time (LSST) will provide a large brightness-complete sample of TNOs for additional followup photometry.
In particular, the discovery of low-$i$ 4:3, 3:2, 2:1, 5:2, and distant resonators would be an excellent sample to search for FaintIR surfaces.
The distribution of FaintIR and BrightIR surfaces in the different Neptune resonances provides useful constraints on the original extent and surface distribution within the proto-planetesimal disk.

\hspace{1cm}
%\begin{acknowledgements}

The authors acknowledge the sacred nature of Maunakea, and appreciate the opportunity to observe from the mountain. 
Based on observations obtained at the international Gemini Observatory, a program of NSF’s NOIRLab, acquired through the Gemini Science Archive and the Gemini Observatory Archive at NSF’s NOIRLab, and processed using the Gemini IRAF package, which is managed by the Association of Universities for Research in Astronomy (AURA) under a cooperative agreement with the National Science Foundation on behalf of the Gemini Observatory partnership: the National Science Foundation (United States), National Research Council (Canada), Agencia Nacional de Investigaci\'{o}n y Desarrollo (Chile), Ministerio de Ciencia, Tecnolog\'{i}a e Innovaci\'{o}n (Argentina), Minist\'{e}rio da Ci\^{e}ncia, Tecnologia, Inova\c{c}\~{o}es e Comunica\c{c}\~{o}es (Brazil), and Korea Astronomy and Space Science Institute (Republic of Korea).
This research used the Canadian Advanced Network For Astronomy Research (CANFAR) operated in partnership by the  Canadian Astronomy Data Centre and The Digital Research Alliance of Canada with support from the National Research Council of Canada, the Canadian Space Agency, CANARIE, and the Canadian Foundation for Innovation.
REP and KV acknowledge funding from NASA Emerging Worlds grant 80NSSC21K0376; KV additionally acknowledges NASA Solar System Workings grant 80NSSC19K0785.
MES was supported by the Science and Technology Facilities Council (STFC) grant ST/V000691/1.
LEB acknowledges funding from the Science Technology Facilities Council (STFC) Grant Code ST/T506369/1.
NP acknowledges funding from Funda\c{c}\~ao para a Ci\^{e}ncia e a Tecnologia (FCT), Portugal, through the research grants UIDB/04434/2020 and UIDP/04434/2020.
Data Access: Data supporting this study are included within the article and the Gemini Observatory Archive (\url{https://archive.gemini.edu}).

%\end{acknowledgements}

\facilities{Gemini} 
\software{TRIPPy \citep{fraser2016phot}}


\begin{thebibliography}{}
\expandafter\ifx\csname natexlab\endcsname\relax\def\natexlab#1{#1}\fi
\providecommand{\url}[1]{\href{#1}{#1}}
\providecommand{\dodoi}[1]{doi:~\href{http://doi.org/#1}{\nolinkurl{#1}}}
\providecommand{\doeprint}[1]{\href{http://ascl.net/#1}{\nolinkurl{http://ascl.net/#1}}}
\providecommand{\doarXiv}[1]{\href{https://arxiv.org/abs/#1}{\nolinkurl{https://arxiv.org/abs/#1}}}

\bibitem[{{Alexandersen} {et~al.}(2016){Alexandersen}, {Gladman}, {Kavelaars},
  {Petit}, {Gwyn}, {Shankman}, \& {Pike}}]{alexandersen2016}
{Alexandersen}, M., {Gladman}, B., {Kavelaars}, J.~J., {et~al.} 2016, \aj, 152,
  111, \dodoi{10.3847/0004-6256/152/5/111}

\bibitem[{{Ali-Dib} {et~al.}(2021){Ali-Dib}, {Marsset}, {Wong}, \&
  {Dbouk}}]{alidib2021}
{Ali-Dib}, M., {Marsset}, M., {Wong}, W.-C., \& {Dbouk}, R. 2021, \aj, 162, 19,
  \dodoi{10.3847/1538-3881/abf6ca}

\bibitem[{{Anderson} \& {Darling}(1952)}]{AndersonDarling}
{Anderson}, T.~W., \& {Darling}, D.~A. 1952, Ann. Math. Statist., 23, 193,
  \dodoi{10.1214/aoms/1177729437}

\bibitem[{{Bannister} {et~al.}(2016){Bannister}, {Kavelaars}, {Petit},
  {Gladman}, {Gwyn}, {Chen}, {Volk}, {Alexandersen}, {Benecchi}, {Delsanti},
  {Fraser}, {Granvik}, {Grundy}, {Guilbert-Lepoutre}, {Hestroffer}, {Ip},
  {Jakubik}, {Jones}, {Kaib}, {Kavelaars}, {Lacerda}, {Lawler}, {Lehner},
  {Lin}, {Lister}, {Lykawka}, {Monty}, {Marsset}, {Murray-Clay}, {Noll},
  {Parker}, {Pike}, {Rousselot}, {Rusk}, {Schwamb}, {Shankman}, {Sicardy},
  {Vernazza}, \& {Wang}}]{bannister2016}
{Bannister}, M.~T., {Kavelaars}, J.~J., {Petit}, J.-M., {et~al.} 2016, \aj,
  152, 70, \dodoi{10.3847/0004-6256/152/3/70}

\bibitem[{{Bannister} {et~al.}(2018){Bannister}, {Gladman}, {Kavelaars},
  {Petit}, {Volk}, {Chen}, {Alexandersen}, {Gwyn}, {Schwamb}, {Ashton},
  {Benecchi}, {Cabral}, {Dawson}, {Delsanti}, {Fraser}, {Granvik},
  {Greenstreet}, {Guilbert-Lepoutre}, {Ip}, {Jakubik}, {Jones}, {Kaib},
  {Lacerda}, {Van Laerhoven}, {Lawler}, {Lehner}, {Lin}, {Lykawka}, {Marsset},
  {Murray-Clay}, {Pike}, {Rousselot}, {Shankman}, {Thirouin}, {Vernazza}, \&
  {Wang}}]{bannister2018}
{Bannister}, M.~T., {Gladman}, B.~J., {Kavelaars}, J.~J., {et~al.} 2018, \apjs,
  236, 18, \dodoi{10.3847/1538-4365/aab77a}

\bibitem[{Barucci {et~al.}(2011)Barucci, Alvarez-Candal, Merlin, Belskaya, {de
  Bergh}, Perna, DeMeo, \& Fornasier}]{barucci2011}
Barucci, M., Alvarez-Candal, A., Merlin, F., {et~al.} 2011, Icarus, 214, 297,
  \dodoi{https://doi.org/10.1016/j.icarus.2011.04.019}

\bibitem[{{Brown} {et~al.}(2011){Brown}, {Burgasser}, \& {Fraser}}]{brown2011}
{Brown}, M.~E., {Burgasser}, A.~J., \& {Fraser}, W.~C. 2011, \apjl, 738, L26,
  \dodoi{10.1088/2041-8205/738/2/L26}

\bibitem[{{Buchanan} {et~al.}(2022){Buchanan}, {Schwamb}, {Fraser},
  {Bannister}, {Marsset}, {Pike}, {Nesvorn{\'y}}, {Kavelaars}, {Benecchi},
  {Lehner}, {Wang}, {Peixinho}, {Volk}, {Alexandersen}, {Chen}, {Gladman},
  {Gwyn}, \& {Petit}}]{buchanan2022}
{Buchanan}, L.~E., {Schwamb}, M.~E., {Fraser}, W.~C., {et~al.} 2022, \psj, 3,
  9, \dodoi{10.3847/PSJ/ac42c9}

\bibitem[Chen et al.(2019)]{chen2019} Chen, Y.-T., Gladman, B., Volk, K., et al.\ 2019, \aj, 158, 214. doi:10.3847/1538-3881/ab480b

\bibitem[{{Chiang} \& {Choi}(2008)}]{Chiang2008}
{Chiang}, E., \& {Choi}, H. 2008, \aj, 136, 350,
  \dodoi{10.1088/0004-6256/136/1/350}

\bibitem[Crompvoets et al.(2022)]{crompvoets2022} Crompvoets, B.~L., Lawler, S.~M., Volk, K., et al.\ 2022, \psj, 3, 113. doi:10.3847/PSJ/ac67e0

\bibitem[{{Dalle Ore} {et~al.}(2013){Dalle Ore}, {Dalle Ore}, {Roush},
  {Cruikshank}, {Emery}, {Pinilla-Alonso}, \& {Marzo}}]{dalleore2013}
{Dalle Ore}, C.~M., {Dalle Ore}, L.~V., {Roush}, T.~L., {et~al.} 2013, \icarus,
  222, 307, \dodoi{10.1016/j.icarus.2012.11.015}

\bibitem[{{Fornasier} {et~al.}(2009){Fornasier}, {Barucci}, {de Bergh},
  {Alvarez-Candal}, {DeMeo}, {Merlin}, {Perna}, {Guilbert}, {Delsanti},
  {Dotto}, \& {Doressoundiram}}]{fornasier2009}
{Fornasier}, S., {Barucci}, M.~A., {de Bergh}, C., {et~al.} 2009, \aap, 508,
  457, \dodoi{10.1051/0004-6361/200912582}

\bibitem[{{Fraser} {et~al.}(2016){Fraser}, {Alexandersen}, {Schwamb},
  {Marsset}, {Pike}, {Kavelaars}, {Bannister}, {Benecchi}, \&
  {Delsanti}}]{fraser2016phot}
{Fraser}, W., {Alexandersen}, M., {Schwamb}, M.~E., {et~al.} 2016, \aj, 151,
  158, \dodoi{10.3847/0004-6256/151/6/158}

\bibitem[{{Fraser} \& {Brown}(2012)}]{fraser2012}
{Fraser}, W.~C., \& {Brown}, M.~E. 2012, \apj, 749, 33,
  \dodoi{10.1088/0004-637X/749/1/33}

  \bibitem[Fraser et al.(2014)]{fraser2014} Fraser, W.~C., Brown, M.~E., Morbidelli, A., et al.\ 2014, \apj, 782, 100. doi:10.1088/0004-637X/782/2/100

\bibitem[Fraser et al.(2023)]{fraser2021} Fraser, W.~C., Pike, R.~E., Marsset, M., et al.\ 2023, \psj, 4, 80. doi:10.3847/PSJ/acc844

\bibitem[{{Gladman} {et~al.}(2008){Gladman}, {Marsden}, \&
  {Vanlaerhoven}}]{gladman2008}
{Gladman}, B., {Marsden}, B.~G., \& {Vanlaerhoven}, C. 2008, {Nomenclature in
  the Outer Solar System} (The University of Arizona Press), 43--57

\bibitem[{{Gladman} \& {Volk}(2021)}]{Gladman2021}
{Gladman}, B., \& {Volk}, K. 2021, \araa, 59,
  \dodoi{10.1146/annurev-astro-120920-010005}

\bibitem[{{Gomes} {et~al.}(2005){Gomes}, {Levison}, {Tsiganis}, \&
  {Morbidelli}}]{Gomes2005}
{Gomes}, R., {Levison}, H.~F., {Tsiganis}, K., \& {Morbidelli}, A. 2005, \nat,
  435, 466, \dodoi{10.1038/nature03676}

\bibitem[{{Gomes}(2003)}]{gomes2003}
{Gomes}, R.~S. 2003, \icarus, 161, 404, \dodoi{10.1016/S0019-1035(02)00056-8}

\bibitem[Greenstreet et al.(2019)]{greenstreet2019} Greenstreet, S., Gladman, B., McKinnon, W.~B., et al.\ 2019, \apjl, 872, L5. doi:10.3847/2041-8213/ab01db

\bibitem[{Hartigan \& Hartigan(1985)}]{hartigan1985}
Hartigan, J.~A., \& Hartigan, P.~M. 1985, The Annals of Statistics, 13, 70 ,
  \dodoi{10.1214/aos/1176346577}

\bibitem[{{Hodapp} {et~al.}(2003){Hodapp}, {Jensen}, {Irwin}, {Yamada},
  {Chung}, {Fletcher}, {Robertson}, {Hora}, {Simons}, {Mays}, {Nolan}, {Bec},
  {Merrill}, \& {Fowler}}]{hodapp2003}
{Hodapp}, K.~W., {Jensen}, J.~B., {Irwin}, E.~M., {et~al.} 2003, \pasp, 115,
  1388, \dodoi{10.1086/379669}

\bibitem[{{Hook} {et~al.}(2004){Hook}, {J{\o}rgensen}, {Allington-Smith},
  {Davies}, {Metcalfe}, {Murowinski}, \& {Crampton}}]{hook2004}
{Hook}, I.~M., {J{\o}rgensen}, I., {Allington-Smith}, J.~R., {et~al.} 2004,
  \pasp, 116, 425, \dodoi{10.1086/383624}

\newblock \doarXiv{2202.09045}
\bibitem[Huang et al.(2022)]{huang2022} Huang, Y., Gladman, B., \& Volk, K.\ 2022, \apjs, 259, 54. doi:10.3847/1538-4365/ac559a


\bibitem[{{Jewitt}(2009)}]{Jewitt2009}
{Jewitt}, D. 2009, \aj, 137, 4296, \dodoi{10.1088/0004-6256/137/5/4296}

\bibitem[Jewitt(2018)]{jewitt2018} Jewitt, D.\ 2018, \aj, 155, 56. doi:10.3847/1538-3881/aaa1a4

\bibitem[Kaib et al.(2011a)]{kaib2011_ICARUS} Kaib, N.~A., Ro{\v{s}}kar, R., \& Quinn, T.\ 2011, \icarus, 215, 491. doi:10.1016/j.icarus.2011.07.037

\bibitem[Kaib et al.(2011b)]{kaib2011_ApJ} Kaib, N.~A., Quinn, T., \& Brasser, R.\ 2011, \aj, 141, 3. doi:10.1088/0004-6256/141/1/3

\bibitem[Kavelaars et al.(2021)]{kavelaars2021} Kavelaars, J.~J., Petit, J.-M., Gladman, B., et al.\ 2021, \apjl, 920, L28. doi:10.3847/2041-8213/ac2c72

\bibitem[{{Lawler} {et~al.}(2018){Lawler}, {Kavelaars}, {Alexandersen},
  {Bannister}, {Gladman}, {Petit}, \& {Shankman}}]{lawlerFASS}
{Lawler}, S.~M., {Kavelaars}, J.~J., {Alexandersen}, M., {et~al.} 2018,
  Frontiers in Astronomy and Space Sciences, 5, 14,
  \dodoi{10.3389/fspas.2018.00014}

\bibitem[{{Levison} {et~al.}(2008){Levison}, {Morbidelli}, {Van Laerhoven},
  {Gomes}, \& {Tsiganis}}]{levison2008}
{Levison}, H.~F., {Morbidelli}, A., {Van Laerhoven}, C., {Gomes}, R., \&
  {Tsiganis}, K. 2008, \icarus, 196, 258, \dodoi{10.1016/j.icarus.2007.11.035}

\bibitem[{{Lykawka} \& {Mukai}(2007)}]{lykawkamukai07}
{Lykawka}, P.~S., \& {Mukai}, T. 2007, Icarus, 192, 238,
  \dodoi{10.1016/j.icarus.2007.06.007}

\bibitem[{{Malhotra}(1995)}]{malhotra1995}
{Malhotra}, R. 1995, \aj, 110, 420, \dodoi{10.1086/117532}

\bibitem[Marsset et al.(2023)]{marsset2022} Marsset, M., Fraser, W.~C., Schwamb, M.~E., et al.\ 2023, \psj, 4, 160. doi:10.3847/PSJ/ace7d0

\bibitem[{{Marsset} {et~al.}(2019){Marsset}, {Fraser}, {Pike}, {Bannister},
  {Schwamb}, {Volk}, {Kavelaars}, {Alexandersen}, {Chen}, {Gladman}, {Gwyn},
  {Lehner}, {Peixinho}, {Petit}, \& {Wang}}]{marsset2019}
{Marsset}, M., {Fraser}, W.~C., {Pike}, R.~E., {et~al.} 2019, \aj, 157, 94,
  \dodoi{10.3847/1538-3881/aaf72e}

\bibitem[{{Murray-Clay} \& {Chiang}(2006)}]{murray-clay2006}
{Murray-Clay}, R.~A., \& {Chiang}, E.~I. 2006, \apj, 651, 1194,
  \dodoi{10.1086/507514}

\bibitem[{{Nesvorn{\'y}}(2015{\natexlab{a}})}]{nesvorny2015b}
{Nesvorn{\'y}}, D. 2015{\natexlab{a}}, \aj, 150, 73,
  \dodoi{10.1088/0004-6256/150/3/73}

\bibitem[{{Nesvorn{\'y}}(2015{\natexlab{b}})}]{nesvorny2015a}
---. 2015{\natexlab{b}}, \aj, 150, 68, \dodoi{10.1088/0004-6256/150/3/68}

\bibitem[{{Nesvorn{\'y}}(2018)}]{Nesvorny2018}
---. 2018, \araa, 56, 137, \dodoi{10.1146/annurev-astro-081817-052028}

\bibitem[{{Nesvorn{\'y}} \& {Morbidelli}(2012)}]{nesvorny_morbidelli2012}
{Nesvorn{\'y}}, D., \& {Morbidelli}, A. 2012, \aj, 144, 117,
  \dodoi{10.1088/0004-6256/144/4/117}

\bibitem[{{Nesvorn{\'y}} {et~al.}(2020){Nesvorn{\'y}}, {Vokrouhlick{\'y}},
  {Alexandersen}, {Bannister}, {Buchanan}, {Chen}, {Gladman}, {Gwyn},
  {Kavelaars}, {Petit}, {Schwamb}, \& {Volk}}]{nesvorny2020}
{Nesvorn{\'y}}, D., {Vokrouhlick{\'y}}, D., {Alexandersen}, M., {et~al.} 2020,
  \aj, 160, 46, \dodoi{10.3847/1538-3881/ab98fb}

\bibitem[{{Ofek}(2012)}]{ofek2012}
{Ofek}, E.~O. 2012, \apj, 749, 10, \dodoi{10.1088/0004-637X/749/1/10}

\bibitem[{{Peixinho} {et~al.}(2015){Peixinho}, {Delsanti}, \&
  {Doressoundiram}}]{peixinho2015}
{Peixinho}, N., {Delsanti}, A., \& {Doressoundiram}, A. 2015, \aap, 577, A35,
  \dodoi{10.1051/0004-6361/201425436}

\bibitem[{{Peixinho} {et~al.}(2012){Peixinho}, {Delsanti}, {Guilbert-Lepoutre},
  {Gafeira}, \& {Lacerda}}]{peixinho}
{Peixinho}, N., {Delsanti}, A., {Guilbert-Lepoutre}, A., {Gafeira}, R., \&
  {Lacerda}, P. 2012, \aap, 546, A86, \dodoi{10.1051/0004-6361/201219057}

\bibitem[{{Petit} {et~al.}(2011){Petit}, {Kavelaars}, {Gladman}, {Jones},
  {Parker}, {Van Laerhoven}, {Nicholson}, {Mars}, {Rousselot}, {Mousis},
  {Marsden}, {Bieryla}, {Taylor}, {Ashby}, {Benavidez}, {Campo Bagatin}, \&
  {Bernabeu}}]{cfeps}
{Petit}, J.-M., {Kavelaars}, J.~J., {Gladman}, B.~J., {et~al.} 2011, \aj, 142,
  131, \dodoi{10.1088/0004-6256/142/4/131}

\bibitem[{{Petit} {et~al.}(2017){Petit}, {Kavelaars}, {Gladman}, {Jones},
  {Parker}, {Bieryla}, {Van Laerhoven}, {Pike}, {Nicholson}, {Ashby}, \&
  {Lawler}}]{hilat}
---. 2017, \aj, 153, 236, \dodoi{10.3847/1538-3881/aa6aa5}

\bibitem[Petit et al.(2023)]{petit2023} Petit, J.-M., Gladman, B., Kavelaars, J.~J., et al.\ 2023, \apjl, 947, L4, \dodoi{doi:10.3847/2041-8213/acc525}

\bibitem[{{Pike} {et~al.}(2017{\natexlab{a}}){Pike}, {Lawler}, {Brasser},
  {Shankman}, {Alexandersen}, \& {Kavelaars}}]{pike2017}
{Pike}, R.~E., {Lawler}, S., {Brasser}, R., {et~al.} 2017{\natexlab{a}}, \aj,
  153, 127, \dodoi{10.3847/1538-3881/aa5be9}

\bibitem[{{Pike} {et~al.}(2017{\natexlab{b}}){Pike}, {Fraser}, {Schwamb},
  {Kavelaars}, {Marsset}, {Bannister}, {Lehner}, {Wang}, {Alexandersen},
  {Chen}, {Gladman}, {Gwyn}, {Petit}, \& {Volk}}]{pike2017z}
{Pike}, R.~E., {Fraser}, W.~C., {Schwamb}, M.~E., {et~al.} 2017{\natexlab{b}},
  \aj, 154, 101, \dodoi{10.3847/1538-3881/aa83b1}

\bibitem[{{Pirani} {et~al.}(2021){Pirani}, {Johansen}, \&
  {Mustill}}]{pirani2021}
{Pirani}, S., {Johansen}, A., \& {Mustill}, A.~J. 2021, \aap, 650, A161,
  \dodoi{10.1051/0004-6361/202037465}

\bibitem[{{Schaller} \& {Brown}(2007)}]{schaller2007}
{Schaller}, E.~L., \& {Brown}, M.~E. 2007, \apjl, 659, L61,
  \dodoi{10.1086/516709}

\bibitem[{{Schwamb} {et~al.}(2019){Schwamb}, {Fraser}, {Bannister}, {Marsset},
  {Pike}, {Kavelaars}, {Benecchi}, {Lehner}, {Wang}, {Thirouin}, {Delsanti},
  {Peixinho}, {Volk}, {Alexandersen}, {Chen}, {Gladman}, {Gwyn}, \&
  {Petit}}]{schwamb2019}
{Schwamb}, M.~E., {Fraser}, W.~C., {Bannister}, M.~T., {et~al.} 2019, \apjs,
  243, 12, \dodoi{10.3847/1538-4365/ab2194}

\bibitem[{{Seccull} {et~al.}(2021){Seccull}, {Fraser}, \&
  {Puzia}}]{seccull2021}
{Seccull}, T., {Fraser}, W.~C., \& {Puzia}, T.~H. 2021, \psj, 2, 57,
  \dodoi{10.3847/PSJ/abe4d9}

\bibitem[{{Sheppard}(2012)}]{Sheppard2012}
{Sheppard}, S.~S. 2012, \aj, 144, 169, \dodoi{10.1088/0004-6256/144/6/169}

\bibitem[{{Van Laerhoven} {et~al.}(2019){Van Laerhoven}, {Gladman}, {Volk},
  {Kavelaars}, {Petit}, {Bannister}, {Alexandersen}, {Chen}, \&
  {Gwyn}}]{vanLaerhoven2019}
{Van Laerhoven}, C., {Gladman}, B., {Volk}, K., {et~al.} 2019, \aj, 158, 49,
  \dodoi{10.3847/1538-3881/ab24e1}

\bibitem[{{Vilenius} {et~al.}(2018){Vilenius}, {Stansberry}, {Muller},
  {Mueller}, {Kiss}, {Santos-Sanz}, {Mommert}, {Pal}, {Lellouch}, {Ortiz},
  {Peixinho}, {Thirouin}, {Lykawka}, {Horner}, {Duffard}, {Fornasier}, \&
  {Delsanti}}]{vilenius2018}
{Vilenius}, E., {Stansberry}, E., {Muller}, E., {et~al.} 2018, \aap,
  \dodoi{https://doi.org/10.1051/0004-6361/201732564}

\bibitem[{{Volk} \& {Malhotra}(2019)}]{volkmalhotra2019}
{Volk}, K., \& {Malhotra}, R. 2019, \aj, 158, 64,
  \dodoi{10.3847/1538-3881/ab2639}

\bibitem[{{Volk} {et~al.}(2016){Volk}, {Murray-Clay}, {Gladman}, {Lawler},
  {Bannister}, {Kavelaars}, {Petit}, {Gwyn}, {Alexandersen}, {Chen}, {Lykawka},
  {Ip}, \& {Lin}}]{volk2016}
{Volk}, K., {Murray-Clay}, R., {Gladman}, B., {et~al.} 2016, \aj, 152, 23,
  \dodoi{10.3847/0004-6256/152/1/23}

\bibitem[{{Volk} {et~al.}(2018){Volk}, {Murray-Clay}, {Gladman}, {Lawler},
  {Yu}, {Alexandersen}, {Bannister}, {Chen}, {Dawson}, {Greenstreet}, {Gwyn},
  {Kavelaars}, {Lin}, {Lykawka}, \& {Petit}}]{Volk2018}
{Volk}, K., {Murray-Clay}, R.~A., {Gladman}, B.~J., {et~al.} 2018, \aj, 155,
  260, \dodoi{10.3847/1538-3881/aac268}

\bibitem[{{Yu} {et~al.}(2018){Yu}, {Murray-Clay}, \& {Volk}}]{Yu2018}
{Yu}, T. Y.~M., {Murray-Clay}, R., \& {Volk}, K. 2018, \aj, 156, 33,
  \dodoi{10.3847/1538-3881/aac6cd}

\end{thebibliography}
\end{document}